%% file: IEEE_TCOM_version_4_onecol.tex
\documentclass[lettersize,journal]{IEEEtran}
\normalsize

\ifCLASSINFOpdf
\else
\fi

\usepackage{graphicx}
\usepackage{epstopdf}
\usepackage[caption=false,font=footnotesize]{subfig}
\usepackage{cite}
\usepackage{color}
\usepackage{xcolor}
\usepackage[cmex10]{amsmath}
\usepackage{amssymb}
\usepackage{threeparttable}
\usepackage{mathtools}
\usepackage{hyperref}
\usepackage{algpseudocode}
\usepackage{multirow}
\usepackage[T1]{fontenc}
\usepackage[ruled]{algorithm2e}
\usepackage{booktabs}
\usepackage{stfloats}
\usepackage{tensor}
\usepackage{bbm}
\usepackage{multicol}
\usepackage{multirow}
\usepackage{tikz}
\usepackage{cooltooltips}

\definecolor{mygreen}{rgb}{0.1,.6,0.1}
 
\DeclareMathOperator*{\argminB}{argmin}   
 \usepackage{multirow}
\usepackage[cmex10]{amsmath} 
\usepackage{makecell}
\setcellgapes{2pt}
\usepackage[thinlines]{easytable}
\newcommand{\floor}[1]{\left\lfloor #1 \right\rfloor}

\def\BibTeX{{\rm B\kern-.05em{\sc i\kern-.025em b}\kern-.08em
    T\kern-.1667em\lower.7ex\hbox{E}\kern-.125emX}}
\ifCLASSOPTIONcompsoc
    \usepackage[caption=false, font=normalsize, labelfont=sf, textfont=sf]{subfig}
\else
\usepackage[caption=false, font=footnotesize]{subfig}
\fi
\usepackage{lipsum}
\usepackage{xcolor}
\definecolor{mygreen}{rgb}{0.1,.6,0.1}

\begin{document}
\title{LDPC Decoding with Degree-Specific Neural Message Weights and RCQ Decoding}

\author{
	\IEEEauthorblockN{ Linfang Wang, Caleb Terrill, Richard Wesel, and  Dariush Divsalar}\\
	\thanks{Part of the contents related to Note-Degree-Based weight sharing and Neural-2D-MinSum decoder in this paper have been published to IEEE ISTC 2021 \cite{wang2021ISTC}.}
	\thanks{Linfang Wang,  Caleb Terrill, Richard Wesel are with the Department of Electrical and Computer Engineering, University of California, Los Angeles, Los Angeles, CA, 90095 USA.  E-mail: \{lfwang, cterrill26, wesel\}@ucla.edu.}
\thanks{Dariush Divsalar is with Jet Propulsion Laboratory, California Institute of Technology, Pasadena, California 91109 E-mail: Dariush.Divsalar@jpl.nasa.gov.

This work was carried out in part at the Jet Propulsion Laboratory, California Institute of Technology, under a contract with the National Aeronautics and Space Administration, and National Science Foundation (NSF) grants CCF-2008918 and CCF-1911166 at UCLA. Any opinions, findings, and conclusions or recommendations expressed in this material are those of the authors and do not necessarily reflect views of NSF.}%
}

%
%

\markboth{IEEE Transactions on Communications}%
{}
\maketitle

\begin{abstract}
Recently, neural networks have improved MinSum message-passing decoders for low-density parity-check (LDPC) codes by multiplying or adding weights to the messages, where the weights are determined by a neural network.  The neural network complexity to determine distinct weights for each edge is high, often limiting the application to relatively short LDPC codes.  Furthermore, storing separate weights for every edge and every iteration can be a burden for hardware implementations.  To reduce neural network complexity and storage requirements, this paper proposes a family of weight-sharing schemes that use the same weight for edges that have the same check node degree and/or variable node degree. Our simulation results show that node-degree-based weight-sharing can deliver the same performance requiring distinct weights for each node.

This paper also combines these degree-specific neural weights with a reconstruction-computation-quantization (RCQ) decoder to produce a weighted RCQ (W-RCQ) decoder. The W-RCQ decoder with node-degree-based weight sharing has a reduced hardware requirement compared with the original RCQ decoder.  As an additional contribution, this paper identifies and resolves a gradient explosion issue that can arise when  training neural LDPC decoders.


\end{abstract}

%
\IEEEpeerreviewmaketitle

\section{Introduction}

\IEEEPARstart{L}{ow}-Density Parity-Check  (LDPC) codes \cite{GallagerPhD1963} have been implemented broadly, including in NAND flash systems and wireless communication systems. Message-passing decoders are often used to decode LDPC codes. Typical message-passing decoders utilize belief propagation (BP), MinSum, and its variations. However, message-passing decoders are sub-optimal because of the existence of cycles in the corresponding Tanner graph.

Recently, numerous works have been focused on enhancing the performance of message-passing decoders with the help of neural networks (NNs) \cite{Nachmani2016-bs,Lugosch2017-ed,Nachmani2017-qq,Nachmani2018-ra,Liang2018-lw,Wu2018-zr,Lugosch2018-gu,Lyu2018-nz,Xiao2019-kj,Deng2019-cf,Abotabl2019-wt,Buchberger2020-pf,Wang2020-fb,Lian2019-jh,nguyen2021neural}, such as neural belief propagation (N-BP) in \cite{Nachmani2016-bs}, normalized MinSum (NMS) and neural OMS decoders in \cite{Nachmani2018-ra, Lugosch2017-ed,Nachmani2016-bs}. The neural network is created by unfolding the message passing operations of each decoding iteration \cite{Nachmani2016-bs}.
Each decoding iteration is unfolded into two hidden layers, representing a check node processing layer and a variable node processing layer, and each neuron represents a variable-to-check message (V2C) or a check-to-variable (C2V) message.  These neural decoders normally assign each C2V message and/or each V2C message a distinct weight in each iteration and hence are impractical for long-blocklength LDPC codes because the number of required parameters is proportional to the number of edges in the Tanner graph of the parity check matrix.

One solution is to share the weights across iterations or edges in the Tanner graph, like in \cite{Nachmani2017-qq,Wang2020-fb,Abotabl2019-wt, Lian2019-jh}. However, these simple weight-sharing methods sacrifice decoding performance in different ways.
Besides, the precursor works of literature mainly \textcolor{blue}{focus} on the short-blocklength codes ($n<2000$), which may have resulted from the fact that the required memory for training neural decoders with long block lengths by using popular deep learning research platforms, such as PyTorch, exceeds the computation resources that researchers can access. 
However, as shown in \cite{Abotabl2019-wt,wang2021ISTC}, it is possible to train neural decoders by only using CPUs on personal computers for very long-blocklength codes if resources are handled more efficiently. 

On the other perspective, decoders for LDPC codes with low message bit widths are desired when considering the limited hardware resources. 
Recently, the non-uniformly quantized decoders  \cite{Planjery2012-FAID,Xiao2020-sb, Lewandowsky2018-IB,Stark2018-IBMA,Stark2019-IBPBRL,Lee2005-RFQThorpe,He2019-RFQCai, Wang2020-kh,terrill2021fpga,wang2022TCOMRCQ}  have shown to deliver excellent performance with very low message precision. One promising decoding paradigm is called reconstruction-computation-quantization (RCQ) decoder \cite{Wang2020-kh,terrill2021fpga,wang2022TCOMRCQ}. 
 The node operation in an RCQ decoder involves a reconstruction function that allows high-precision message computation and a quantization function that allows low-precision message passing between nodes. Specifically, the reconstruction function, equivalent to a dequantizer,  maps the low-bitwidth messages received by a node to  high-bitwidth messages for computation. The quantization function quantizes the calculated high-bitwidth messages to low-bitwidth messages that will be sent to its neighbor nodes. 

 The excellent decoding performance of RCQ decoder comes from its dynamic quantizers and dequantizers that are updated in each layer and each iteration. 
 However, such dynamic quantizers/dequantizers are also \textcolor{blue}{overheads} of the RCQ decoder in hardware implementation, which may even offset the benefit brought by the low bit-width messages\cite{wang2022TCOMRCQ}.


\subsection{Contribution}
   \textcolor{black}{This paper proposes a family of weight-sharing schemes for the neural MinSum decoder based on the check node degree and variable node degree. Our simulation results show that the decoders with the node-degree-based weight-sharing schemes can deliver the same performance as the neural MinSum decoder that doesn't share the weights. This paper also combines neural decoding with the RCQ decoding paradigm and proposes a weighted RCQ (W-RCQ) decoder. The W-RCQ decoder with node-degree-based weight sharing has a reduced hardware requirement compared with the RCQ decoder. The contributions of this paper are summarized below:} 
\begin{itemize}
    \item \textit{Posterior Joint Training Method.} This paper identifies the gradient explosion issue when training neural LDPC decoders. A posterior joint training method is proposed in this paper to address the gradient explosion problem. Simulation results show posterior joint training delivers better decoding performance than the simple gradient clipping method.
     \item \textit{Node-Degree-Based Weight Sharing}. 
     This paper illustrates that the weight values of the N-NMS decoder are strongly related to check node degree, variable node degree, and iteration index. As a result, this paper proposes node-degree-based weight-sharing schemes that assign the same weight to the edges with the \textcolor{black}{same check node degree and/or variable node degree}.
     \item \textit{Neural-2D-MinSum decoder}. By employing the node-degree-based weight-sharing schemes on the N-NMS and N-OMS decoders, this paper proposes the N-2D-NMS decoder and N-2D-OMS decoder. \emph{2D} means 2-dimensional and implies that the weights in each iteration are shared \textcolor{blue}{across} two dimensions, i.e., check node degree and variable node degree.
    \item  \textit{W-RCQ Decoder.} This paper applies N-2D-NMS and N-2D-OMS to the RCQ decoding paradigm to introduce a weighted-RCQ (W-RCQ) decoding paradigm.  Simulation results for a (9472,8192) LDPC code on a field-programmable gate array (FPGA) device show that compared with the 4-bit RCQ decoder and the 5-bit OMS decoder,  the 4-bit W-RCQ decoder delivers comparable FER performance but with reduced hardware requirements.
\end{itemize}

\subsection{Organization}

The remainder of this paper is organized as follows: Section \ref{sec: effi_NNMS} derives the gradients for a flooding-scheduled N-NMS decoder and shows that the memory to calculate the gradients can be reduced by storing the forward messages compactly. This section also describes the posterior joint training method that addresses the gradient explosion issue.
Section \ref{sec: N-2D-NMS} proposes  node-degree-based weight-sharing schemes for neural MinSum decoder, which leads to a family of neural-2D-MinSum decoders.
Section \ref{sec: W-RCQ} gives the W-RCQ decoding structure and describes how to train W-RCQ  parameters via a \textcolor{blue}{quantized neural network (QNN)}. The simulation results are presented in Section \ref{sec: Simulation}, and Section \ref{sec: conclusion} concludes our work. 

\section{Training Neural MinSum Decoders for Long Blocklength Codes}\label{sec: effi_NNMS}
For the neural network corresponding to a neural LDPC decoder, the number of neurons in each hidden layer equals the number of edges in the Tanner graph corresponding to the parity check matrix  \cite{Nachmani2016-bs}. For the popular NN platforms, such as PyTorch, each neuron requires a data structure that stores the value of the neuron, the gradient of the neuron, the connection of this neuron with other neurons, and so on. Therefore, \textcolor{black}{training} neural decoders for long-blocklength LDPC codes in PyTorch \textcolor{blue}{demands} significant memory, which poses a challenge for researchers with limited resources.

However, the data structure used in PyTorch is redundant to the neural LDPC decoders. One reason is that the neuron connections between hidden layers are repetitive and can be interpreted by the parity check matrix.
This immediately reduces the required memory.  
This section uses N-NMS decoder to show that the memory required to calculate gradients of the neural MinSum decoders can be further reduced by compactly storing the messages in forward propagation.

\subsection{Forward Propagation of N-NMS Decoder}
Let $H\in\mathbb{F}_2^{(n-k)\times n}$ be the parity check matrix of an $(n,k)$ binary LDPC code, where $n$ is the codeword length and $k$ is dataword length. Denote $i^{th}$ variable node and $j^{th}$ check node by $v_i$ and $c_j$, respectively. 
\textcolor{blue}{Let $\mathrm{sgn}(\cdot)$ be the sign function, i.e., $\mathrm{sgn}(x)=1$ for $x\geq0$ and $\mathrm{sgn}(x)=-1$ otherwise.}
For the flooding-scheduled decoder, in the $t^{th}$ decoding iteration, N-NMS decoder updates the C2V message, $u^{(t)}_{c_j \rightarrow v_i}$, by: 
\begin{align}
\begin{split}
     u^{(t)}_{c_i\rightarrow v_j} &= \beta^{(t)}_{(c_i,v_j)} \times  \prod_{v_{j'}\in \mathcal{N}(c_i)\setminus\{v_j\}} \text{sgn}\left(l^{(t-1)}_{v_{j'}\rightarrow c_{i}}\right) \\ &\times  \min_{v_{j'}\in \mathcal{N}(c_i)\setminus\{v_j\}} \left|l^{(t-1)}_{v_{j'}\rightarrow c_{i}}\right|\label{equ: flooding_1},
\end{split}
\end{align}
\begin{figure}[t]
    \centering
    \includegraphics[width=0.7\linewidth]{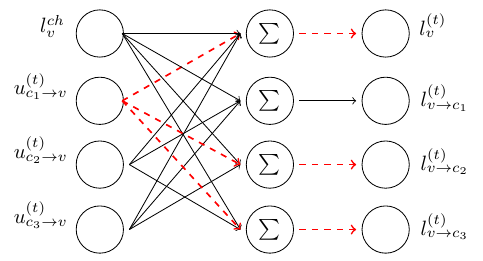}
    \caption{Computation of V2C messages of a degree-3 variable node $v$. The red dashed paths show that the gradient of the node $u^{(t)}_{c_i \rightarrow v_j}$  comes from the nodes $l^{(t)}_{v}$, $l^{(t)}_{v\rightarrow c_2}$ and $l^{(t)}_{v\rightarrow c_3}$.}
    \label{fig:v2c_calculation}
\end{figure}

$\mathcal{N}(c_i)$ is the set of variable nodes that connect $c_i$ and 
$\bigl\{\beta^{(t)}_{(c_i,v_j)}|i\in\{1,\ldots (n-k)\},j\in\{1,\ldots n\}, \allowbreak H(i,j)=1, t\in\{1,\ldots, I_T\} \bigr\}$ is the set of  trainable parameters. $I_T$ represents the maximum iterations. The V2C message, $l^{(t)}_{v_i\rightarrow c_j}$, and posterior of each variable node, $l_{v_i}^{(t)}$, of N-NMS decoder in iteration $t$
 are calculated by:
\begin{align}
l^{(t)}_{v_j\rightarrow c_i} &=  l^{ch}_{v_j} + \sum_{c_{i'}\in \mathcal{N}(v_j)\setminus\{c_i\}} u^{(t)}_{c_{i'}\rightarrow v_j}, \label{equ: v2c_update}\\
l^{(t)}_{v_j} &= l^{ch}_{v_j} + \sum_{c_{i'}\in \mathcal{N}(v_j)} u^{(t)}_{c_{i'}\rightarrow v_j}.\label{equ: pos_update}
\end{align}
$\mathcal{N}(v_j)$ represents the set of the check nodes connected to $v_j$. $l^{ch}_{v_j}$ is the log-likelihood ratio (LLR) of the channel observation of $v_j$. The decoding stops when all \textcolor{blue}{parity check nodes} are satisfied or $I_T$ is reached.

\begin{figure*}[ht]
    \centering
    \includegraphics[width=0.7\linewidth]{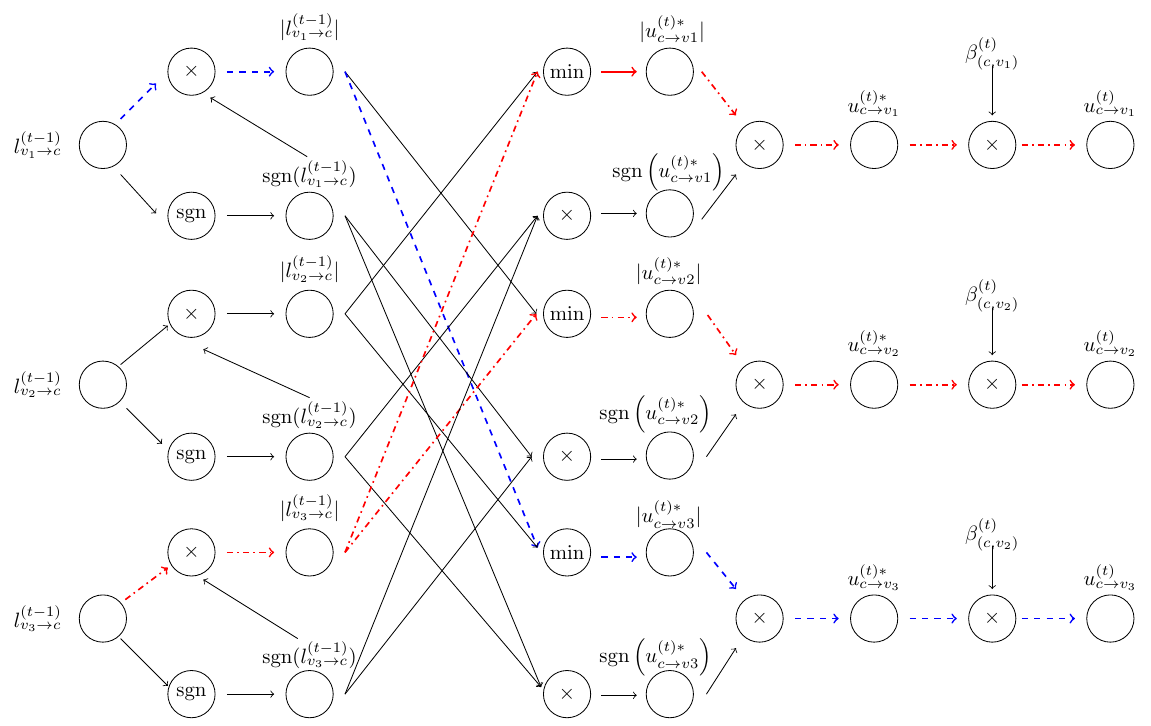}
    \caption{Computation of V2C messages of a degree-3 check node $c$. The example assumes that $v_3$ and $v_1$ provide the first and second minimum values, respectively. As a result, only $l^{(t-1)}_{v_1\rightarrow c}$ and $l^{(t-1)}_{v_3\rightarrow c}$ accumulate the gradients, which are illustrated as the blue dashed and red dash-dotted paths, respectively.}
    \label{fig:c2v_diag}
\end{figure*}
\subsection{Backward Propagation of N-NMS}
Let $J$ be some loss function for the N-NMS neural network, for example, the multi-loss cross entropy in \cite{Nachmani2016-bs}. Denote the gradients of loss $J$ with respect to (w.r.t.) the trainable weights, the C2V message and V2C message by $\frac{\partial J}{ \partial \beta^{(t)}_{(c_i,v_j)}}$, $\frac{\partial J}{\partial u^{(t)}_{c_i\rightarrow v_j}}$, and $\frac{\partial J}{\partial l^{(t)}_{v_j \rightarrow c_i}}$ respectively.

\textcolor{black}{ Fig. \ref{fig:v2c_calculation} shows the calculation of the V2C messages  for a degree-3 variable node $v$ that connects check nodes $c_1$, $c_2$, and $c_3$. The gradients of these V2C messages are components of the gradients of the C2V messages. As shown by the red dashed paths,   $u^{(t)}_{c_1\rightarrow v}$ is used to calculate $l^{(t)}_{v\rightarrow c_2}$, $l^{(t)}_{v\rightarrow c_3}$, and $l^{(t)}_v$, therefore the gradient $\frac{\partial J}{u^{(t)}_{c_1\rightarrow v}}$ is accumulated by these three terms. Generally,}  in iteration $t$, $\frac{\partial J}{\partial u^{(t)}_{c_i \rightarrow v_j}}$ is updated by:
\begin{align}
\label{equ: update-c}
    \frac{\partial J}{\partial u^{(t)}_{c_i \rightarrow v_j}} = \frac{\partial J}{\partial l^{(t)}_{v_j}}+ \sum_{c_{i'}\in \mathcal{N}(v_j)\setminus\{c_i\}}\frac{\partial J}{\partial l^{(t)}_{ v_j\rightarrow c_{i'} }}.
\end{align}

In iteration $t$, when calculating C2V messages, check node $c_i$ receives minimum and second minimum magnitude values, denoted by $\texttt{min1}^{t}_{c_i}$ and \textcolor{blue}{ $\texttt{min2}^{t}_{c_i}$} from variable nodes $\texttt{pos1}^{t}_{c_i}$ and $\texttt{pos2}^{t}_{c_i}$, respectively, where
\begin{align}
\begin{split}
    \texttt{min1}^{t}_{c_i} &= \min_{v_{j'}\in \mathcal{N}(c_i)} |l^{(\textcolor{black}{t-1})}_{v_{j'}\rightarrow c_{i}}|, \\
     \texttt{pos1}^{t}_{c_i} &= \argminB_{v_{j'}\in \mathcal{N}(c_i)} |l^{(\textcolor{black}{t-1})}_{v_{j'}\rightarrow c_{i}}|.\label{equ: min1}
\end{split}\\
\begin{split}
    \texttt{min2}^{t}_{c_i} &= \min_{v_{j'}\in \mathcal{N}(c_i)/\{\texttt{pos1}^{t}_{c_i}\}}|l^{(\textcolor{black}{t-1})}_{v_{j'}\rightarrow c_{i}}|,\\
     \texttt{pos2}^{t}_{c_i} &= \argminB_ {v_{j'}\in \mathcal{N}(c_i)/\{\texttt{pos1}^{t}_{c_i}\}}|l^{(\textcolor{black}{t-1})}_{v_{j'}\rightarrow c_{i}}|\label{equ: pos}.
\end{split}
\end{align}

\textcolor{black}{Only $\texttt{min1}^{t}_{c_i}$ and $\texttt{min2}^{t}_{c_i}$ are used for C2V messages calculation. Fig. \ref{fig:c2v_diag} illustrates an example of computing the C2V messages of a degree-3 check node $c$ that connects variable nodes $v_1$, $v_2$, and $v_3$. Fig. \ref{fig:c2v_diag} assumes $\texttt{pos1}^{(t)}_{c}=v_3$ and \textcolor{blue}{$\texttt{pos2}^{(t)}_{c}=v_1$}. }

\textcolor{black}{
It can be seen from Fig. \ref{fig:c2v_diag} that  $\frac{\partial J}{\partial \beta^{(t)}_{(c_i,v_j)}}$ is calculated using $\frac{\partial J}{\partial u^{(t)}_{c_i \rightarrow v_j}}$ by:
\begin{align}\label{equ: gradient_beta}
    \frac{\partial J}{\partial \beta^{(t)}_{(c_i,v_j)}} = u^{(t)*}_{c_i\rightarrow v_j}\frac{\partial J}{\partial u^{(t)}_{c_i\rightarrow v_j}},
\end{align}}
where $u^{(t)*}_{c_i\rightarrow v_j}=\frac{u^{(t)}_{c_i \rightarrow v_j}}{\beta^{(t)}_{(c_i, v_j)}}$. $u^{(t)*}_{c_i\rightarrow v_j}$ is the output of check node Min operation and hence can be calculated efficiently by ${\rm{sgn}}\left(l^{(t-1)}_{v_j\rightarrow c_i}\right)$, $\texttt{min1}^{t}_{c_i}$, $\texttt{min2}^{t}_{c_i}$, $\texttt{pos1}^{t}_{c_i}$.

\textcolor{black}{In Fig. \ref{fig:c2v_diag}, $l^{(t-1)}_{v_3\rightarrow c}$ is used to compute $u^{(t)}_{c\rightarrow v_1}$ and $u^{(t)}_{c\rightarrow v_2}$, $l^{(t-1)}_{v_1\rightarrow c}$ is used to compute \textcolor{blue}{$u^{(t)}_{c\rightarrow v_3}$}.   $l^{(t-1)}_{v_2\rightarrow c}$ is not involved in computing any C2V messages. As a result, only $l^{(t-1)}_{v_1\rightarrow c}$ and $l^{(t-1)}_{v_3\rightarrow c}$ accumulate the gradients, as shown in the blue dashed and red dash-dotted paths in Fig. \ref{fig:c2v_diag}, respectively. Generally, for all variable nodes connected to the check node $c_i$,   only $l^{(t-1)}_{c_i\rightarrow \texttt{pos1}^{(t)}_{c_i}}$ and $l^{(t-1)}_{c_i\rightarrow \texttt{pos2}^{(t)}_{c_i}}$ receive backward information. Note that the sign operation makes gradient 0, and $\mathrm{min}$ operation passes the gradient to the neuron that provides the minimum value.    Hence, $\frac{\partial J}{\partial l^{(t-1)}_{v_j\rightarrow c_i}}$ is computed as follows:
}
   \begin{align}
    \left\{ \begin{array}{l l} \text{sgn}\left(l^{(t-1)}_{v_j\rightarrow c_i}\right)\sum_{v_{j'}\in \mathcal{N}
    (c_i)\setminus \{v_j\}} \frac{\partial J}{\partial |{u^{(t)*}_{c_i \rightarrow v_{j'}}} |} & , v_j = \texttt{pos1}^{(t)}_{c_i}   \\ \text{sgn}\left(l^{(t-1)}_{v_j\rightarrow c_i}\right) \frac{\partial J}{\partial \left|{u^{(t)*}_{ c_i \rightarrow  \texttt{pos1}^{(t)}_{c_i}}} \right|}  &  , v_j = \texttt{pos2}^{(t)}_{c_i} \\  0 & , \text{otherwise}  . \\ \end{array} \right. 
\end{align} 
The term $\frac{\partial J}{\partial |{u^{(t)*}_{c_i \rightarrow v_j}} |}$ is calculated by:
\begin{align}\label{equ: end_back_prob}
    \frac{\partial J}{\partial |{u^{(t)*}_{c_i \rightarrow v_j}} |}= \text{sgn}(u^{(t)*}_{c_i \rightarrow v_j}) \beta_{(c_i,v_j)}^{(t)}\frac{\partial J}{\partial {u^{(t)}_{c_i \rightarrow v_j}} }.
\end{align}




\eqref{equ: update-c}-\eqref{equ: end_back_prob} indicate that the neuron values in each hidden layer can be stored compactly with sgn$\left(l^{(t)}_{v_j \rightarrow c_i}\right)$, $\texttt{min1}^{t}_{c_i}$, $\texttt{min2}^{t}_{c_i}$, $\texttt{pos1}^{t}_{c_i}$ and $\texttt{pos1}^{t}_{c_i}$. The compactly stored neural values in the hidden layers \textcolor{black}{significantly reduce memory}.  

\subsection{Posterior Joint Training}\label{Sec: gradient_explosion}
\begin{figure}[t] 
    \centering
  \subfloat[\label{fig: gradient_explosion}]{%
      \includegraphics[width=1.0\linewidth]{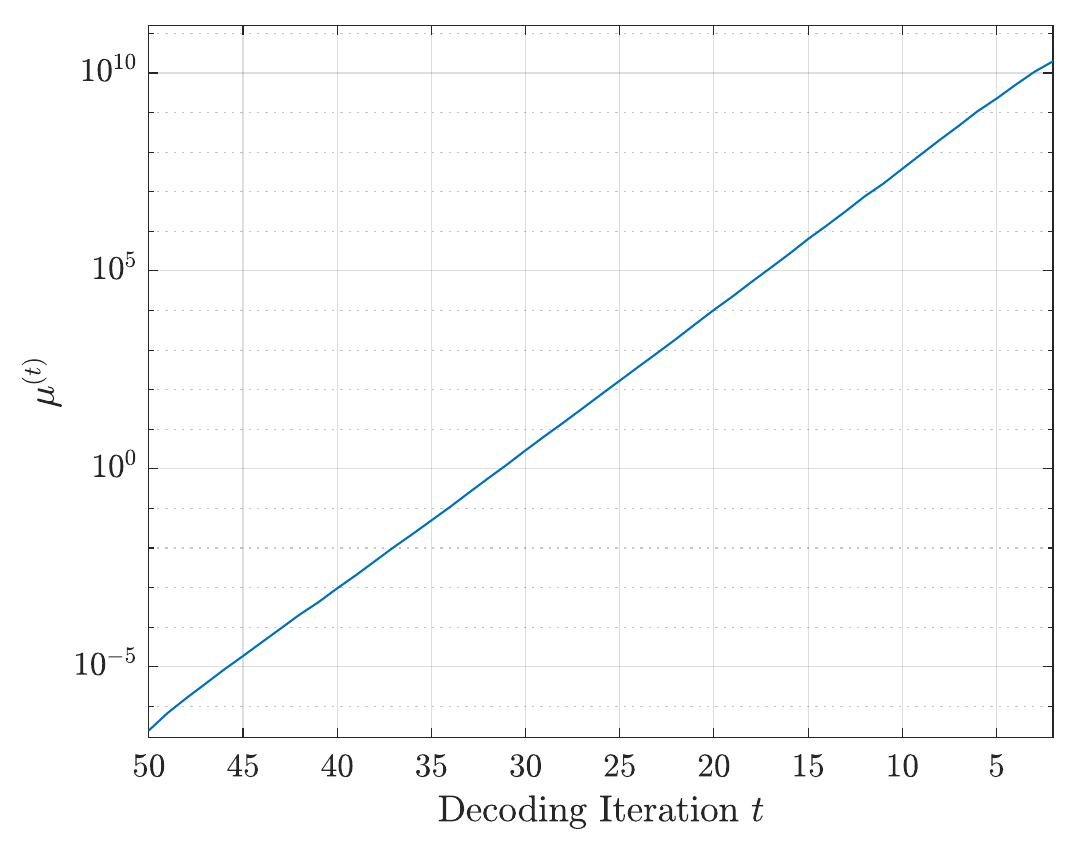}}
 \\
  \subfloat[\label{fig: FER-FNNMS-1}]{%
        \includegraphics[width=1.0\linewidth]{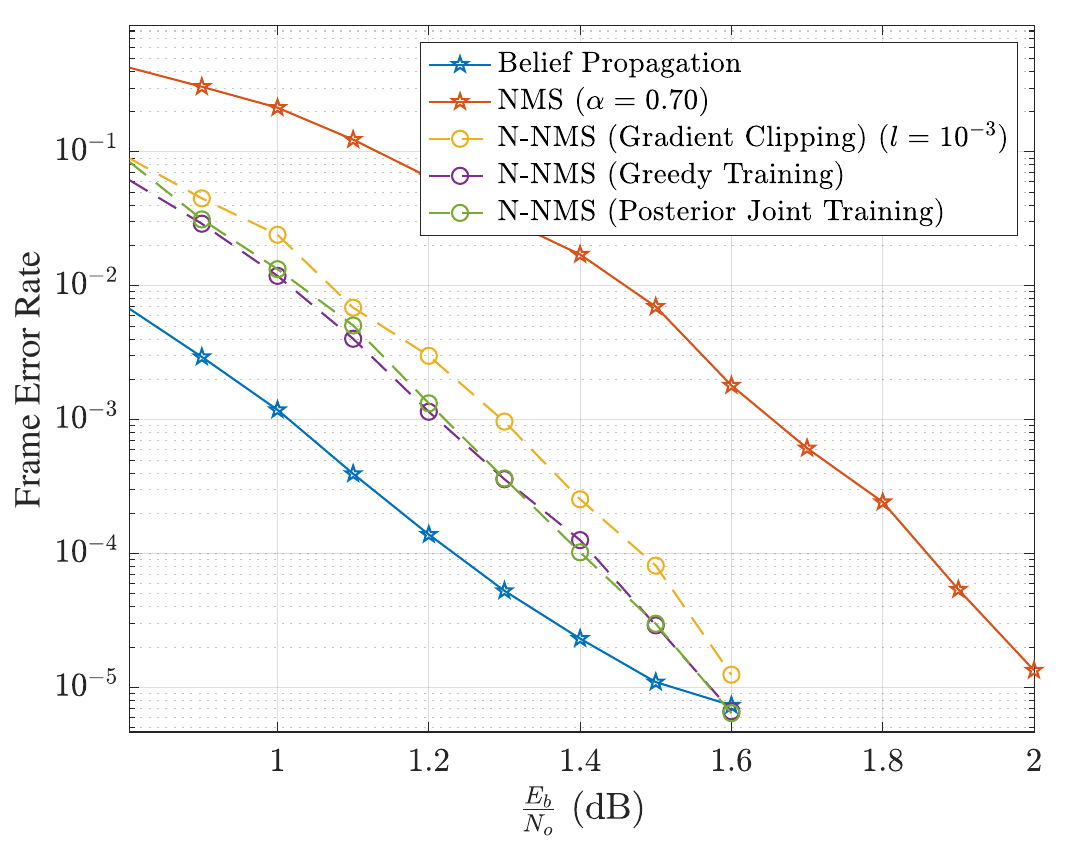}}
  \caption{Fig. (a): The average magnitude of gradients of loss $J$ w.r.t. C2V messages in each decoding iteration. 
  Fig. (b): FER curves of the flooding-scheduled N-NMS decoders for a (3096,1032) LDPC code.  Gradient clipping, greedy training, and posterior joint training are used to address the gradient explosion issue. }
\end{figure}
Eq. \eqref{equ: l_update} implies that in iteration $t$, for all variable nodes that connect check node $c$, only $\texttt{pos1}_{c}^{t}$ and $\texttt{pos2}_{c}^{t}$ receive gradients from $c$. Besides, $|\mathcal{N}(c)|-1$ gradient terms flow to $\texttt{pos1}_{c}^1$. Hence, if check node  $c$ has a large degree, the gradient of $J$ w.r.t. $\texttt{pos1}_{c}^{t}$ can have a large magnitude, and this large-magnitude gradient will be propagated to the neurons in the preceding layer corresponded to the C2V messages whose check nodes (other than $c$) connect $\texttt{pos1}_{c}^{t}$.  As a result, the large-magnitude gradients are accumulated and propagated as back propagation proceeds, which results in gradient explosion.

Fig. \ref{fig: gradient_explosion} illustrates the gradient explosion phenomenon when training a flooding-scheduled N-NMS decoder for a (3096,1032) LDPC code. Define $\mu^{(t)}$ as the average magnitude of the gradients of $J$ w.r.t. all C2V messages in iteration $t$. The gradients are calculated by feeding the NN corresponding to  N-NMS decoder with a random input sample and then performing backward propagation. Fig. \ref{fig: gradient_explosion} plots $\mu^{(t)}$ in each decoding iteration. The maximum check node degree and variable node degree of the code are 19 and 27, respectively. \textcolor{blue}{The maximum number of decoding iterations} of the decoder is 50.  It can be seen that the $\mu^{(t)}$ increases exponentially with the decrease of decoding iteration $t$.

Eq. \eqref{equ: gradient_beta} indicates that  large magnitude of $\frac{\partial J}{\partial u^{(t)}_{(c,v)}}$ leads to large magnitude of $\frac{\partial J}{\partial \beta^{(t)}_{(c,v)}}$  and hence prevents the neural network from optimizing weights effectively. To our knowledge, this paper is the first to report the gradient explosion issue for neural LDPC decoder training. However, there have been several techniques that solve the gradient explosion problem:

\begin{enumerate} 
    \item \textit{Gradient Clipping}. Gradient explosion is a common problem in the deep learning field, such as recurrent neural networks, and one way to solve this problem is gradient clipping \cite{GoodBengCour16}. 
    The gradient clipping in this paper means to limit the maximum gradient magnitude to be some threshold $l$.
    
    \item \textit{Greedy Training}. Dai \textit{et al} in \cite{dai2021learning} proposed greedy training. Greedy training trains the parameters in $t^{th}$ decoding iteration by fixing the pre-trained parameters in the first $t-1$ iterations. Greedy training solves the gradient explosion problem because the large magnitude gradients won't be accumulated and propagated to the preceding hidden layers.
    However, greedy training requires a time complexity that is proportional to $I_T^2$, because one must have trained first $(t-1)$ iterations in order to train
    the $t^{th}$ decoding iteration.

\end{enumerate}    

Eq. \eqref{equ: update-c} indicates that the gradient of $J$ w.r.t. $u^{(t)}_{c_i\rightarrow v_j}$ comes from two parts: the first part is from the posterior $l^{(t)}_{v_j}$, and the second part is from the V2C messages $l^{(t)}_{v_j\rightarrow c_i'}$, $c_{i'}\in \mathcal{N}(v_j) \setminus \{c_i\}$. Based on the previous analysis, if any $l^{(t)}_{v_j\rightarrow c_i'}$, $c_{i'}\in \mathcal{N}(v_j) \setminus \{c_i\}$ provides a large magnitude gradient, the neuron $u^{(t)}_{c_i\rightarrow v_j}$  can also have a large magnitude gradient. This will result in a large magnitude to the gradient of $J$ w.r.t. $\beta^{(t)}_{(c_i,v_j)}$, as indicated by \eqref{equ: gradient_beta}.  In this paper, we propose posterior joint training, which calculates the gradient of $J$ w.r.t. $u^{(t)}_{c_i\rightarrow v_j}$ only using the posterior $l^{(t)}_{v_j}$.  More explicitly, for the flooding-scheduled  N-NMS neural network, $\frac{\partial J}{\partial u^{(t)}_{c_i\rightarrow v_j}}$ is calculated by:
    \begin{align}\label{equ: pt_flooding}
         \frac{\partial J}{\partial u^{(t)}_{c_i \rightarrow v_j}} = \frac{\partial J}{\partial l^{(t)}_{v_j}}.
    \end{align}
    Hence,  the gradient of $J$ w.r.t. $\beta^{(t)}_{(c_i,v_j)}$ is calculated as:
    \begin{align}\label{equ: gradient_beta_2}
    \frac{\partial J}{\partial \beta^{(t)}_{(c_i,v_j)}} = \frac{u^{(t)}_{c_i \rightarrow v_j}}{\beta^{(t)}_{(c_i,v_j)}} \frac{\partial J}{\partial u^{(t)}_{(c_i,v_j)}}=\frac{u^{(t)}_{c_i \rightarrow v_j}}{\beta^{(t)}_{c_i\rightarrow v_j}}\frac{\partial J}{\partial l^{(t)}_{v_j}}.
    \end{align}
By calculating the gradients of neurons in the $t^{th}$ decoding iteration only using the posterior  $l_{v_j}^{(t)}$, \eqref{equ: pt_flooding} and \eqref{equ: gradient_beta_2} prevent the large-magnitude gradients from being propagated to the preceding hidden layers. 
 \textcolor{black}{The posterior training equivalently treats each decoding iteration as the last iteration, because $\frac{\partial J}{\partial u^{(t)}_{c_i \rightarrow v_j}}$ in the last iteration is calculated by \eqref{equ: pt_flooding}. Besides, \eqref{equ: pt_flooding} is also used in the greedy training \cite{dai2021learning}, because the greedy training trains the parameters of iteration $t$ by assuming the decoder has a maximum iteration of $t$ and fixing the pre-trained parameters of previous $t-1$ iterations.}  However, the posterior joint training optimizes parameters of all decoding iterations jointly and hence requires a time complexity proportional to $I_T$.

\textcolor{black}{Unlike the flooding scheduled decoder, which calculates the V2C message using the C2V messages all from the previous iterations, the layered-scheduled decoder facilitates the most recently updated C2V messages to calculate V2C messages. As a result, the $u^{(t)}_{c_i\rightarrow v_j}$ is used to calculate the following terms: 1) Soft decision in iteration $t$, \textcolor{blue}{$l^{(t)}_{v_j}$}; 2) V2C messages in iteration $t$, $l^{(t)}_{v_j \rightarrow c_{i'}}$, where $i'\in\{i^*|c_{i^*}\in\mathcal{N}(v_j),i^*>i\}$ ; and 3) V2C messages in iteration $t+1$, $l^{(t+1)}_{v_j \rightarrow c_{i'}}$, where $i'\in\{i^*|c_{i^*}\in\mathcal{N}(v_j),i^*<i\}$. Hence, $\frac{\partial J}{\partial u^{(t)}_{c_i \rightarrow v_j}}$ is calculated by: 
\begin{align}\label{equ: conv_layer}
\begin{split}
\frac{\partial J}{\partial u^{(t)}_{c_i \rightarrow v_j}} &=  \frac{\partial J}{\partial l^{(t)}_{v_j}}+ \sum_{i'\in\{i^*|c_{i^*}\in\mathcal{N}(v_j),i'>i\}}\frac{\partial J}{\partial l^{(t)}_{v_j \rightarrow c_{i'}}}\\&+\sum_{i'\in \{i^*|c_{i^*}\in\mathcal{N}(v_j),i'<i\}}\frac{\partial J}{\partial l^{(t+1)}_{v_j \rightarrow c_{i'}}}. 
\end{split}
\end{align} }
    
Posterior joint training abandons the last term in \eqref{equ: conv_layer} and calculates $\frac{\partial J}{\partial u^{(t)}_{c_i\rightarrow v_j}}$ as follows:
    \begin{align}\label{equ: pt_layer}
        \frac{\partial J}{\partial u^{(t)}_{c_i \rightarrow v_j}} =  \frac{\partial J}{\partial l^{(t)}_{v_j}}+ \sum_{i'\in \{i^*|c_{i^*}\in\mathcal{N}(v_j),i^*>i\}}\frac{\partial J}{\partial l^{(t)}_{v_j \rightarrow c_{i'}}}.
    \end{align}

Fig. \ref{fig: FER-FNNMS-1} shows the frame error rate (FER) of flooding-scheduled N-NMS decoders for a (3096,1032) LDPC code. The maximum decoding iteration time is 50. All three methods to prevent gradient explosion are implemented. The gradient clipping uses a threshold of $l=10^{-3}$. The performance of BP and NMS decoders with the same decoding schedule and maximum decoding iteration \textcolor{blue}{is} also compared. The NMS decoder uses a factor of 0.7. The simulation results show that greedy training and posterior joint training have similar FER curves and perform better than gradient clipping. However,  posterior joint training has a lower time complexity than greedy training.

\section{Node-Degree-Based Weight Sharing}\label{sec: N-2D-NMS}

\begin{figure}[t]
    \centering
  \subfloat[\label{fig: mean-t}]{%
       \includegraphics[width=0.8\linewidth]{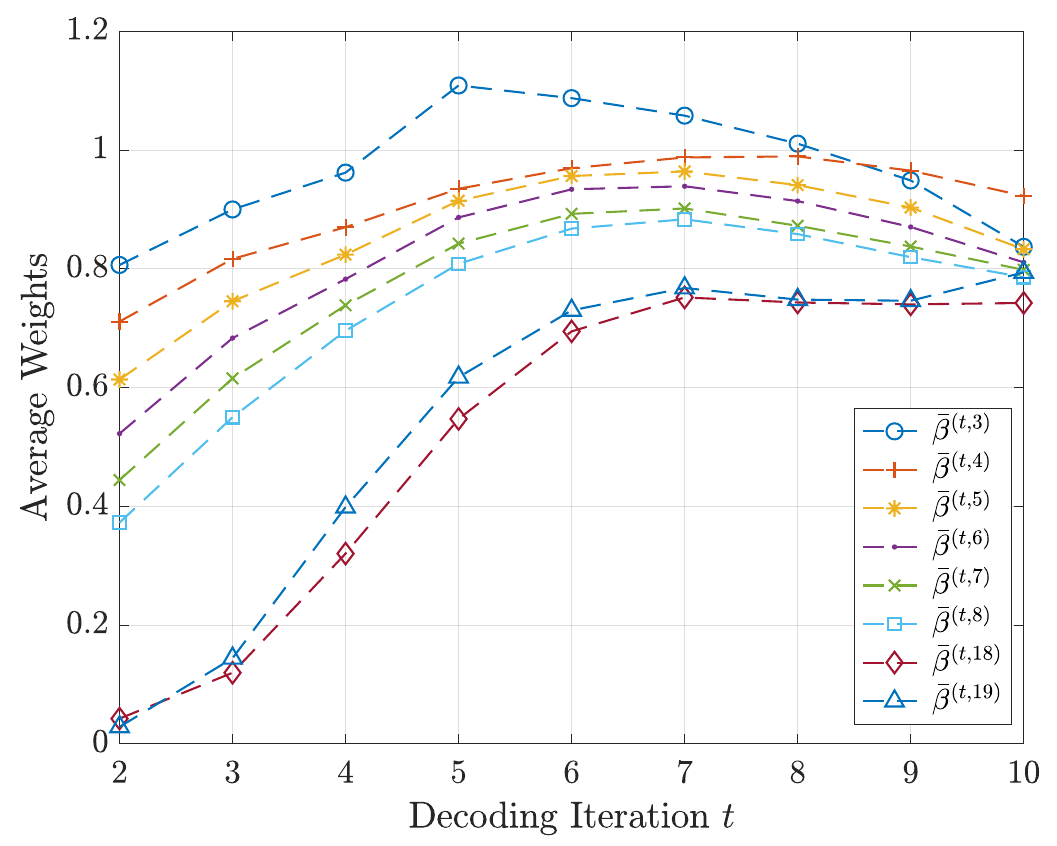}}
\\
  \subfloat[\label{fig: degree-19-iter4}]{%
        \includegraphics[width=0.8\linewidth]{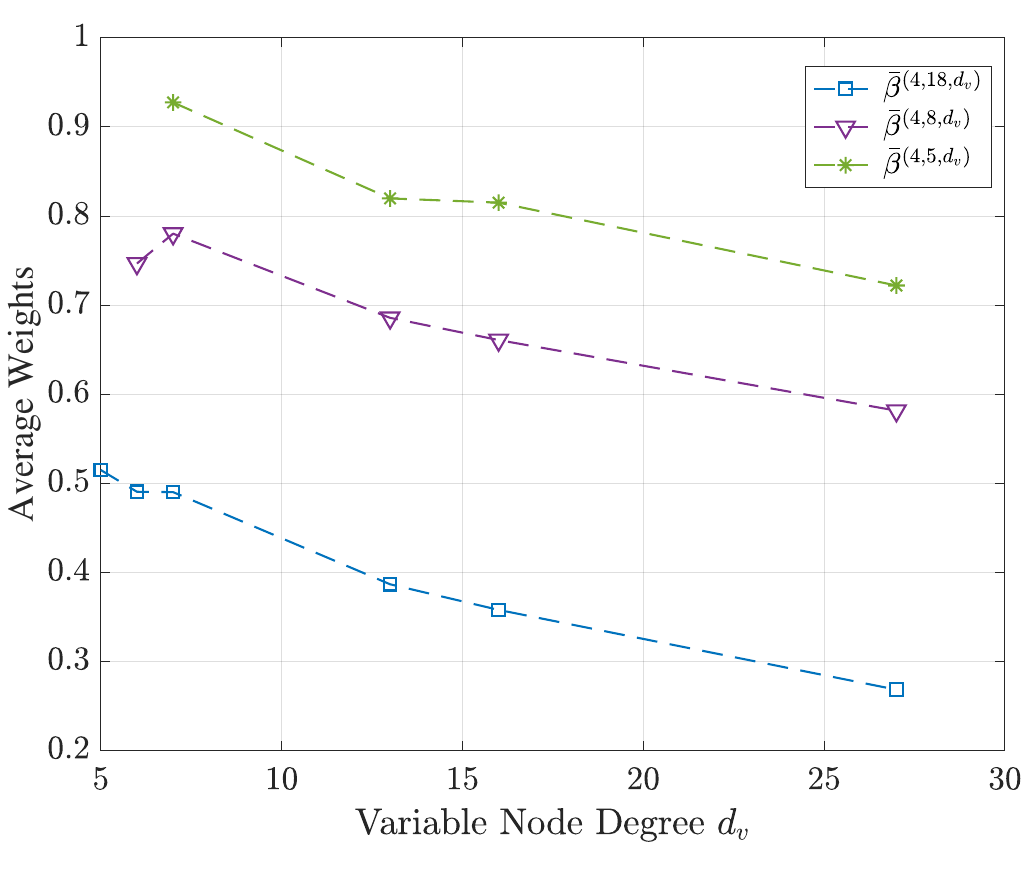}}
  \caption{Mean values of messages of a flooding-scheduled N-NMS decoder for a (3096,1032) LDPC code in each iteration show strong correlations to check node degree and variable node degree.}
   \label{fig: weights_evaluation}
\end{figure}

N-NMS and N-OMS \textcolor{black}{decoders} for the long-blocklength LDPC codes are impractical, because the number of parameters of these decoders is proportional to the number of edges in the corresponding Tanner graph. Weight sharing \cite{xie2021weight} solves this problem by assigning one weight to different neurons in the NN. Different weight-sharing schemes have been proposed to reduce the number of neural weights in N-NMS and N-OMS \textcolor{black}{decoders}. However, simple weight-sharing schemes, such as across iterations or edges in \cite{Wang2020-fb,Lian2019-jh}, degrade the decoding performance in different degrees. 

This section proposes the node-degree-based weight-sharing schemes that assign the same weights to the edges with the same check node degree and/or variable node degree. We call the N-NMS and N-OMS decoder with node-degree-based weight-sharing schemes by neural 2-dimensional NMS (N-2D-NMS) \textcolor{black}{decoder} and neural 2-dimensional OMS (N-2D-OMS) decoder, respectively,  because they are similar to the 2D-MS decoders in \cite{Juntan_Zhang2005-2dnms,2doms}. Simulation results in Section \ref{sec: Simulation} show that the N-2D-NMS decoder can deliver the same decoding performance with the N-NMS decoder.

\subsection{Motivation}
In this subsection, we investigate the relationship between the neural weights of a flooding-scheduled N-NMS decoder and node degrees. 
The N-NMS decoder is trained for a  (3096, 1032) LDPC code, the same one used in Section  \ref{Sec: gradient_explosion}. The maximum decoding iteration is 10.

Define the set of neural weights of N-NMS decoder that are associated to check node degree $d_c$ in the $t^{th}$ decoding iteration by $\mathcal{B}^{(t,d_c)}$, and $\mathcal{B}^{(t,d_c)}=\{\beta^{(t)}_{(c_i,v_j)}|\text{deg}(c_i) = d_c\}$. Let $\bar{\beta}^{(t,d_c)}$ be the mean value of $\mathcal{B}^{(t,d_c)}$. Fig.\ref{fig: mean-t} shows $\bar{\beta}^{(t,d_c)}$ versus decoding iteration $t$ with all possible check node degrees. The simulation result shows a clear relationship between check node degree and $\bar{\beta}^{(t,d_c)}$, \textcolor{black}{i.e.,} a larger check node degree corresponds to a smaller $\bar{\beta}^{(t,d_c)}$. This difference is significant in the first few iterations.
Additionally, $\bar{\beta}^{(t,d_c)}$ changes drastically in first few iterations for all check node degrees. 

In order to explore the relationship between the weights and variable node degrees given a check node degree $d_c$ and decoding iteration index $t$, we further define $ \mathcal{B}^{(t,d_c,d_v)}=\{\beta^{(t)}_{(c_i,v_j)}\allowbreak| \text{deg}(c_i)=d_c,\text{deg}(v_i)=d_v\}$. We denote  the average value of $\mathcal{B}^{(t,d_c,d_v)}$ by $\bar{\beta}^{(t,d_c,d_v)}$. Fig. \ref{fig: degree-19-iter4} gives the average weights corresponding to various check node degrees and variable node degrees at iteration $4$. Statistical results show that, given a specific iteration $t$ and check node degree $d_c$, a larger $d_v$ indicates a smaller $\bar{\beta}^{(t,d_c,d_v)}$. 

In conclusion, the weights of the N-NMS decoder are correlated with the check node degree, the variable node degree, and the decoding iteration index. Thus, node degrees should affect the weighting of messages on their incident edges when decoding LDPC codes. This observation motivates us to propose a family of N-2D-MS decoders. 

\subsection{Neural 2D Normalized MinSum Decoders}
Based on the previous discussion, it is intuitive to consider assigning the same weights to messages with same check node degree and/or variable node degree. 
In this subsection, we propose a family \textcolor{blue}{of} node-degree-based weight-sharing schemes.
These weight-sharing schemes can be used on the  N-NMS decoder, which gives \textcolor{black}{an} N-2D-NMS decoder. 

In the $t^{th}$ iteration, a flooding-scheduled N-2D-NMS decoder update $u^{(t)}_{c_i\rightarrow v_j}$ as follows:
\begin{align}
\begin{split}
    u^{(t)}_{c_i\rightarrow v_j} &= \beta^{(t)}_{*} \times  \prod_{v_{j'}\in \mathcal{N}(c_i)/\{v_j\}} \text{sgn}\left(l^{(t-1)}_{v_{j'}\rightarrow c_{i}}\right)  \\&\times  \min_{v_{j'}\in \mathcal{N}(c_i)/\{v_j\}} \left|l^{(t-1)}_{v_{j'}\rightarrow c_{i}}\right|.
\end{split}
     \end{align}
     \begin{align}
      l^{(t)}_{v_j\rightarrow c_i} &=  l^{ch}_{v_i} + \alpha^{(t)}_{*}  \sum_{c_{i'}\in \mathcal{N}(v_j)/\{c_i\}} u^{(t)}_{c_{i'}\rightarrow v_j},
      \end{align}
      \begin{align}
    l^{(t)}_{v_j} &=  l^{ch}_{v_i} + \alpha^{(t)}_{*}  \sum_{c_{i'}\in \mathcal{N}(v_j)} u^{(t)}_{c_{i'}\rightarrow v_j}.
\end{align}

\begin{center}
    \input{weight_sharing_one_col}
\end{center}

$\beta^{(t)}_{*}$ and $\alpha^{(t)}_{*}$ are the learnable weights. The subscript * is replaced in Table \ref{tab:weight_sharing} with the information needed to identify the specific weight depending on the weight-sharing methodology. 
Table \ref{tab:weight_sharing} lists different weight-sharing types, each identified in the first column by a type number. As a special case, we denote type 0 by assigning distinct weights to each edge, i.e., N-NMS decoder. Columns 2 and 3 describe how each type assigns $\beta^{(t)}_{*}$ and $\alpha^{(t)}_{*}$, respectively. In this paper, we refer to a decoder that uses a type-$x$ weight-sharing scheme as a type-$x$ decoder. 

Types 1-4 assign the same weights based on the node degree. In particular, Type 1 assigns the same weight to the edges that have same check node \emph{and} variable node degree. Type 2 considers the check node degree and variable node degree separately. As a simplification, type 3 and type 4 only consider check node degree and variable node degree, respectively.

Dai. \emph{et. al} in \cite{dai2021learning} studied weight sharing based on the edge type of multi-edge-type (MET)-LDPC codes, or protograph-based codes. We also consider this metric for types 5, 6, and 7. Type 5  assigns the same weight to the edges with the same edge type, i.e., the edges that belong to the same position in the protomatrix. In Table. \ref{tab:weight_sharing}, $f$ is the lifting factor.  Types 6 and 7 assign parameters based only on the horizontal (protomatrix row) and vertical layers (protomatrix column), respectively.
Finally, type 8 assigns a single weight to all edges in each decoding iteration, as in \cite{Lian2019-jh,Abotabl2019-wt}.

\begin{figure}[t]
    \centering
    \includegraphics[width=0.6\linewidth]{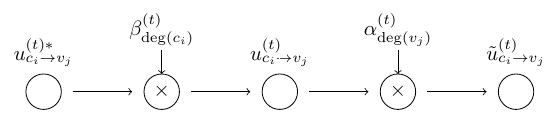}
    \caption{Relationship between $u^{(t)*}_{c_i\rightarrow v_j}$, $u^{(t)}_{c_i\rightarrow v_j}$, and $\tilde{u}^{(t)*}_{c_i\rightarrow v_j}$ in the type-2 N-2D-NMS decoder.} 
    \label{fig: relation}
\end{figure}

\textcolor{black}{
The gradients $\frac{\partial J}{\partial \beta^{(t)}_{*}}$ and $\frac{\partial J}{\partial \alpha^{(t)}_{*}}$ in N-2D-NMS decoder are accumulated from the gradients of C2V messages that use $ \beta^{(t)}_{*}$ and $\alpha^{(t)}_{*}$ in decoding process, respectively. For example, the type-2 N-2D-NMS decoder assigns $\beta^{(t)}_{d_c}$ to all C2V messages with check node degree $d_c$ and assigns $\alpha^{(t)}_{d_v}$ to all C2V messages with check node degree $d_v$. As a result, 
 \begin{align}
    \frac{\partial J}{\partial \beta^{(t)}_{d_c}} &= \sum_{(c_i,v_j)\in\mathcal{E}(d_c)} u^{(t)*}_{c_i\rightarrow v_j} \frac{\partial J}{\partial u^{(t)}_{c_i\rightarrow v_j}},\\
    \frac{\partial J}{\partial \alpha^{(t)}_{d_v}} &= \sum_{(c_i,v_j)\in \mathcal{E}(d_v)}u^{(t)}_{c_i\rightarrow v_j}\frac{\partial J}{\partial \tilde{u}^{(t)}_{c_i\rightarrow v_j}},
 \end{align}
 where $\mathcal{E}(d_c)$ and $\mathcal{E}(d_v)$ are the set of edges whose check node degree and variable node degree are $d_c$ and $d_v$, respectively. $u^{(t)*}_{c_i\rightarrow v_j}=\frac{u^{(t)}_{c_i \rightarrow v_j}}{\beta^{(t)}_{(c_i, v_j)}}$, $\tilde{u}^{(t)}_{c_i\rightarrow v_j}=\alpha^{(t)}_{{\rm{deg}}(v_j)} u^{(t)}_{c_i\rightarrow v_j}$. Fig. \ref{fig: relation} gives the relationship between $u^{(t)*}_{c_i\rightarrow v_j}$, $u^{(t)}_{c_i\rightarrow v_j}$, and $\tilde{u}^{(t)*}_{c_i\rightarrow v_j}$ in the type-2 N-2D-NMS decoder.
} 

A (3096,1032) LDPC code and the (16200,7200) DVBS-2\cite{noauthor_2019-nv} standard LDPC code are considered in this section, and the number of parameters per iteration required for various weight-sharing schemes of these two codes are listed in column 4 and 5 in Table. \ref{tab:weight_sharing}, respectively. It is shown that the number of parameters required by the node-degree-based weight sharing is less than that required by the protomatrix-based weight sharing.

\subsection{Neural 2D Offset MinSum Decoder}
The node-degree-based weight-sharing schemes can be applied to the N-OMS decoder similarly and lead to \textcolor{black}{a} neural 2D OMS (N-2D-OMS) decoder. Specifically, a flooding N-2D-OMS decoder updates  $u^{(t)}_{c_i\rightarrow v_j}$ by: 
\begin{align}
\begin{split}
    u^{(t)}_{c_i\rightarrow v_j} &= \prod_{v_{j'}\in \mathcal{N}(c_i)/\{v_j\}} \text{sgn}\left(l^{(t-1)}_{v_{j'}\rightarrow c_{i}}\right)  \\&\times   \text{ReLu} \left(\min_{v_{j'}\in \mathcal{N}(c_i)/\{v_j\}} \left|(l^{(t-1)}_{v_{j'}\rightarrow c_{i}})\right|-\beta^{(t)}_*-\alpha ^{(t)}_*\right).
\end{split}
\end{align}
${\rm{ReLu}}(x)=\max(0,x)$. The $l_{v_j\rightarrow c_i}^{(t)}$ and $l_{v_j}^{(t)}$ are updated using \eqref{equ: v2c_update} and \eqref{equ: pos_update}. For the N-2D-OMS decoders, the constant value 1 in Table \ref{tab:weight_sharing} should be replaced by 0.

\subsection{Hybrid Neural Decoder}
We consider a hybrid training structure that utilizes a neural network combining feed-forward and recurrent modules to reduce the number of parameters further. The hybrid neural decoder uses distinct neural weights for each of the first $I'$ decoding iterations and uses the same weights 
for the remaining  $I_T-I'$ iterations. 
\textcolor{black}{
For example, for the hybrid neural NMS decoder,  the C2V messages are updated by:
\begin{align}
\label{equ: l_update}
    u^{(t)}_{c_i\rightarrow v_j} = \left\{ \begin{array}{l l} \beta^{(t)}_{(c_i,v_j)} u^{(t)*}_{c_i\rightarrow v_j}, & t<I' \\ \beta^{(I_T)}_{(c_i,v_j)} u^{(t)*}_{c_i\rightarrow v_j},  &   I' \leq t \leq I_T, \\\end{array} \right.
\end{align}
and the hybrid version of N-2D-NMS decoders can be constructed similarly.}

The motivation for the hybrid decoder is from the observation that the neural weights of N-NMS decoder change drastically in the first few iterations but negligibly during the last few iterations, as illustrated in Fig. \ref{fig: weights_evaluation}. Therefore, using the same parameters for the last few iterations doesn't cause large performance degradation.


\section{Weighted RCQ Decoder}\label{sec: W-RCQ}
This section combines the N-2D-NMS or N-2D-OMS decoder with RCQ decoding paradigm and proposes a weighted RCQ (W-RCQ) decoder. Unlike the RCQ decoder, whose quantizers and dequantizers are updated in each iteration (and each layer, if layer-scheduled decoding is considered), the W-RCQ decoder only uses a small number of quantizers and dequantizers during the decoding process. However, the C2V messages of the W-RCQ decoder will be weighted by dynamic node-degree-based parameters that are trained by a QNN.


\subsection{\textcolor{black}{Layered Decoding and RCQ decoder}}
\textcolor{black}{
The flooding schedule and layered schedule are two decoding schedules widely used in LDPC decoders.
As in \eqref{equ: flooding_1} to \eqref{equ: pos_update}, the flooding schedule first updates all C2V messages and then all V2C messages in one iteration. 
The layered schedule, on the other hand, partitions all the check nodes (or variable nodes) to several layers and updates the C2V \textcolor{blue}{messages} and V2C messages layer by layer. 
As an example, in the $t^{th}$ iteration, a layered MinSum decoder calculates the messages $u^{(t)}_{c_i\rightarrow v_{j'}}$ and updates the posteriors $l_{v_{j'}}$ as follows:
\begin{align}
    {l}_{v_{j'}} &=  l_{v_{j'}}- u^{(t-1)}_{c_i\rightarrow v_{j'}}~~\forall v_{j'}\in\mathcal{N}(c_i),\label{equ: v-c}
\end{align}
\begin{align}
\begin{split}
     u^{(t)}_{c_i\rightarrow v_{j'}} &=
    \left(\prod_{v_{\tilde{j}}\in\mathcal{N}(c_i)/\{v_{j'}\}}\textcolor{blue}{\text{sgn}}(l_{v_{\tilde{j}}})\right)\\ & \times\min_{v_{\tilde{j}}\in\mathcal{N}(c_i)/\{v_{j'}\}}|l_{v_{\tilde{j}}}|,~~\forall v_{j'}\in\mathcal{N}(c_i), \label{equ: c-v}
\end{split}
\end{align}
\begin{align}
    l_{v_{j'}} &= {l}_{v_{j'}}+u^{(t)}_{c_i \rightarrow v_{j'}} ~~\forall v_{j'}\in\mathcal{N}(c_i)\label{equ: posteriot_up}.
\end{align}
}

\textcolor{black}{
Low-bit-width decoders with uniform quantizers typically suffer large degradation in decoding performance. The authors in \cite{wang2022TCOMRCQ} propose the reconstruction-computation-quantization (RCQ) paradigm that facilitates dynamic non-uniform quantization to achieve good decoding performance with low message precision. 
}

\begin{figure}[t]
    \centering
  \subfloat[\label{fig: msrcq}]{%
       \includegraphics[width=0.8\linewidth]{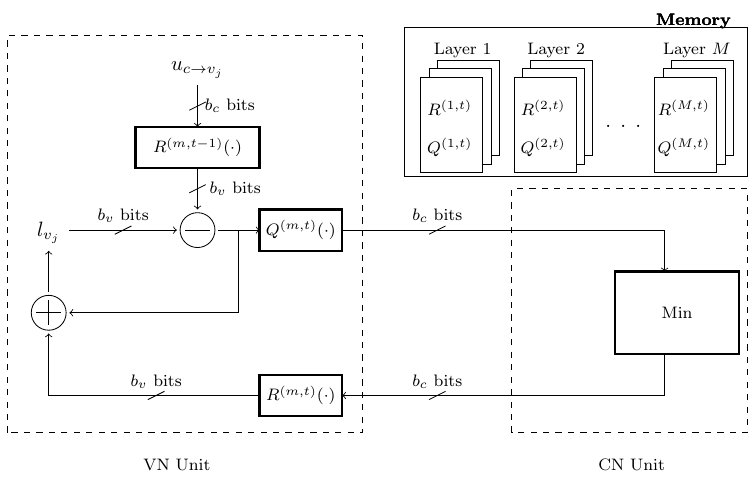}}
   \\
  \subfloat[\label{fig: womsrcq}]{       \includegraphics[width=0.8\linewidth]{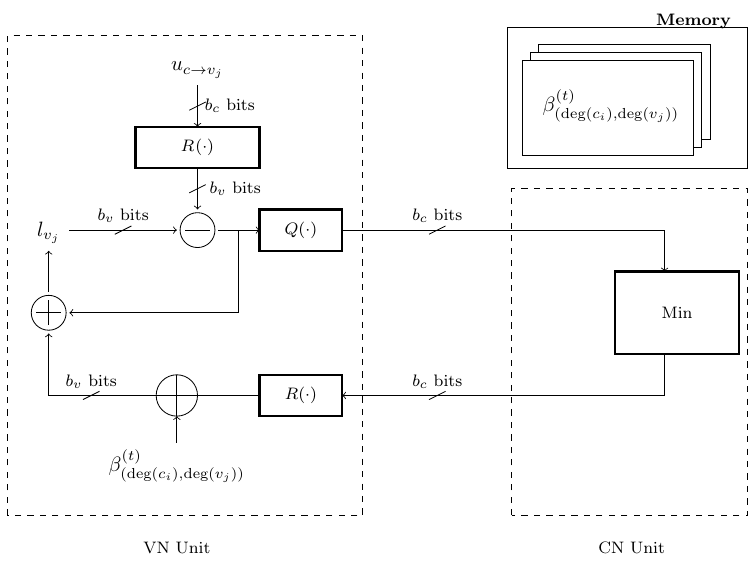}}
  \caption{Decoding diagrams for two distinct versions of layered MinSum decoding: (a) standard RCQ decoder of \cite{wang2022TCOMRCQ} and (b) proposed weighted (offset) RCQ decoder.
}
   \label{fig: rcq_structure}
\end{figure}

\textcolor{black}{
Fig. \ref{fig: msrcq} gives a layered MinSum RCQ (msRCQ) decoding diagram.  The layered msRCQ decoder follows the equation \eqref{equ: v-c}-\eqref{equ: posteriot_up} but with the non-uniform quantizers $Q^{(t,r)}$ and reconstruction functions $R^{(t,r)}$ designed distinctly for each layer  $r=1,\ldots,M$ in iteration $t=1,\ldots, I_T$. $M$ is the number of total layers. The $Q^{(t,r)}$ functions quantize $b_v$-bit messages to $b_c$-bit messages, where $b_v>b_c$, and the $R^{(t,r)}$ functions map $b_c$-bit messages to $b_v$-bit messages.
The dynamic non-uniform $Q^{(t,r)}$ and $R^{(t,r)}$ deliver good decoding performance with a small $b_c$ but also \textcolor{blue}{bring} extra overhead to store those parameters in hardware. As shown in \ref{fig: msrcq}, extra memory is required to store $Q^{(t,r)}$ and   $R^{(t,r)}$ parameters, and extra logic and wires are required to distribute the quantizer 
 and dequantizers to each variable node (VN) units. As shown in \cite{terrill2021fpga,wang2022TCOMRCQ}, the overhead to store a larger number of  $Q^{(t,r)}$  and $R^{(t,r)}$ functions may offset the benefit brought by decoding with messages of low bit width.
}

\subsection{\textcolor{black}{Weighted-RCQ Decoder}}
\textcolor{black}{In this section, we combine the RCQ decoder with the neural decoder to propose a weighted RCQ decoding structure.
Fig. \ref{fig: womsrcq} gives the decoding paradigm of a layer-scheduled weighted OMS-RCQ decoder (W-OMS-RCQ). One goal of W-RCQ is to reduce memory requirements by reducing the number of quantizer/dequantizer pairs  $Q()/R()$ and instead relying on the scalar parameters $\beta^{(t)}_{(\deg(c_i) \deg(v_j))}$ determined by the neural network to capture most or all of the dynamic adjustments needed to adapt with each iteration.  The differences between the W-OMS-RCQ decoder and the msRCQ decoder are summarized as follows: }  
\begin{itemize}
    \item \textit{Reconstruction and Quantization}. 
    \textcolor{black}{Unlike the msRCQ decoding diagram in Fig. \ref{fig: msrcq},} the W-OMS-RCQ decoder only uses very few  $R(\cdot)$ and $Q(\cdot)$ functions in the decoding, for example, three or fewer, 
   and each $R(\cdot)$ and $Q(\cdot)$ are used for several iterations.
   \textcolor{black}{This reduces required memory and wire complexity}.
    \item \textit{Message adjustment}. \textcolor{black}{The} W-OMS-RCQ decoder weights the reconstructed C2V messages with additive parameters. As shown in Fig. \ref{fig: womsrcq}, extra memory is required to store the weights. Besides, the weights in Fig.  \ref{fig: womsrcq} can be multiplicative, leading to a W-NMS-RCQ decoder.
\end{itemize}

\subsection{Non-Uniform Quantizer}
An important design for a W-RCQ decoder is the quantization and reconstruction (dequantization) function selection. The authors in \cite{wang2022TCOMRCQ} use discrete density evolution to design dynamic quantizers and dequantizers for the RCQ decoder. 
For the W-RCQ decoder, this paper considers the quantizers and dequantizers parameterized by the power functions. 

Let $Q(x)$ be a symmetric $b_c$-bit quantizer that features sign information and a magnitude quantizer $Q^*(|x|)$. The magnitude quantizer selects one of $2^{b_c-1}$ possible indices using the threshold values $\{\tau_0,\tau_1,...,\tau_{\text{max}}\}$, where  $\tau_j=C\left(\frac{j}{2^{b_c-1}}\right)^{\gamma}$ for $j\in\{0,...,2^{b_{c}-1}-1\}$ and $\tau_{\text{max}}$ is $\tau_{j_{\text{max}}}$ for $j_{\text{max}} = 2^{b_c-1}-1$.
Given an input $x$, which can be decomposed into sign part ${{\rm{sgn}}(x)}$ and magnitude part $|x|$, $Q^*(|x|)\in \mathbb{F}_2^{b_c-1}$ is defined by:
\begin{align}
   {Q}^*(|x|)=\left\{\begin{matrix}
 j, &    \tau_j\leq|x|<\tau_{j+1}\\
 2^{b_c-1}-1, & |x|\geq \tau_{\mathrm{max}} 
\end{matrix}\right.,
\label{equ: 10}
\end{align}
where $0\leq j\le j_{\text{max}}$. \textcolor{black}{Let $s(x)$ be the sign bit of $x$, which is computed by $s(x)=\mathbbm{1}(x<0)$, where $\mathbbm{1}(\cdot)$ is the indicator function. Then, $Q(x)=[s(x)~Q^*(|x|)]$.}
The thresholds of $Q^*(|x|)$ have a power-function form and are controlled by two parameters. The parameter $C$ defines the maximum magnitude the quantizer can take, and $\gamma$ manipulates the non-uniformity of the quantizer. 
Specifically, if $\gamma=1$,  $Q(x)$ becomes a uniform quantizer.

Let $d\in\mathbb{F}_2^{b_c}$ be a $b_c$-bit message. $d$ can be represented as $[d^{\text{MSB}}\ \tilde{d}]$, where $d^{\text{MSB}}\in\{0,1\}$ indicates the sign and $\Tilde{d}\in\mathbb{F}_2^{b_c-1}$ corresponds to the magnitude. The magnitude reconstruction function ${R}^*(\tilde{d})=\tau_{\tilde{d}} = C\left(\frac{\tilde{d}}{2^{b_c}-1}\right)^{\gamma}$,
and $R(d)=(-2d^{\mathrm{MSB}}+1)R^*(\tilde{d})$. Note that both the magnitude quantization function and magnitude reconstruction function use  $\{\tau_0,...,\tau_{\text{max}}\}$ as their parameters. 

\textcolor{black}{The choice of quantizers is heuristic.  We start with one quantizer and tune its $C$ and $\gamma$. One-quantizer scheme will be used if the resultant decoder has no error floor above a target FER such as $10^{-6}$; otherwise, we increase the number of quantizers by one each time and tune their parameters until a set of quantizers that don't show error floor is found. Besides, for the case of multiple quantizers, the general rule of tuning parameters is that the quantizer for earlier iterations takes a smaller magnitude. }

\textcolor{black}{ The other way to optimize the quantizers is to treat the quantizer parameters as the learnable parameters of the neural network and optimize them in the training process, as in \cite{Xiao2020-sb}. This method will be studied in our future research.}

\subsection{Training W-RCQ decoder via a Quantized Neural Network}
\textcolor{black}{
Like the neural NMS decoder in \cite{Nachmani2016-bs}, the W-RCQ decoder can be unfolded to an NN. The neural network unfolded by the W-RCQ decoder is a QNN because of its quantization and reconstruction functions.  
The QNN then trains the weights of the W-RCQ decoder. 
The training was conducted using stochastic gradient descent with mini-batches.
Each sample in the mini-batch is a BPSK-modulated all-zero codeword corrupted by the additive white Gaussian noise whose variance is in the range where a conventional NMS decoder with a factor 0.7 reaches a FER between $10^{-3}$ and $10^{-2}$. 
In our training experiments, we assign each sample with an $\frac{E_b}{N_0}$  for noise generation.
The $\frac{E_b}{N_0}$ of each sample is chosen such that all the $\frac{E_b}{N_0}$ values with a step of 0.1 dB in the specified range are evenly distributed to each mini-batch. }

One problem of QNN is that quantization functions \textcolor{black}{result} in zeros derivatives almost everywhere.
In this work, we use a straight-through estimator (STE)\cite{Bengio2013-kn, Xiao2020-sb} in the backward propagation \textcolor{black}{to solve this issue}.
\textcolor{black}{
The STE uses artificial gradients in  QNN training to replace the zero derivative of a quantization function in the chain rule. STE is found to be the most efficient training method for QNNs in \cite{Bengio2013-kn}.}

%


\subsection{Fixed-Point W-RCQ decoder}
This paper uses the pair ($b_c$,$b_v$) to denote the bit width for the fixed-point decoders, where $b_c$ is the bit width of C2V messages, and $b_v$ is the bit width of V2C messages and the posteriors of variable nodes. For the W-RCQ decoders,  the learnable parameters are first trained under a floating point message representation and then quantized to $b_v$ bits.

\input{code_info}

\section{Simulation Result and Discussion}\label{sec: Simulation}
This section evaluates the performance of the N-2D-NMS decoders and the W-RCQ decoders for LDPC codes with different block lengths and code rates. The LDPC codes used in this section are listed in Table \ref{tab: code_info}. All the encoded bits are modulated by binary phase-shift keying (BPSK) and transmitted through an Additive White Gaussian Noise (AWGN) channel. 

\begin{figure}[t] 
    \centering
  \subfloat[\label{fig: dvbs2-fer1}]{%
       \includegraphics[width=0.8\linewidth]{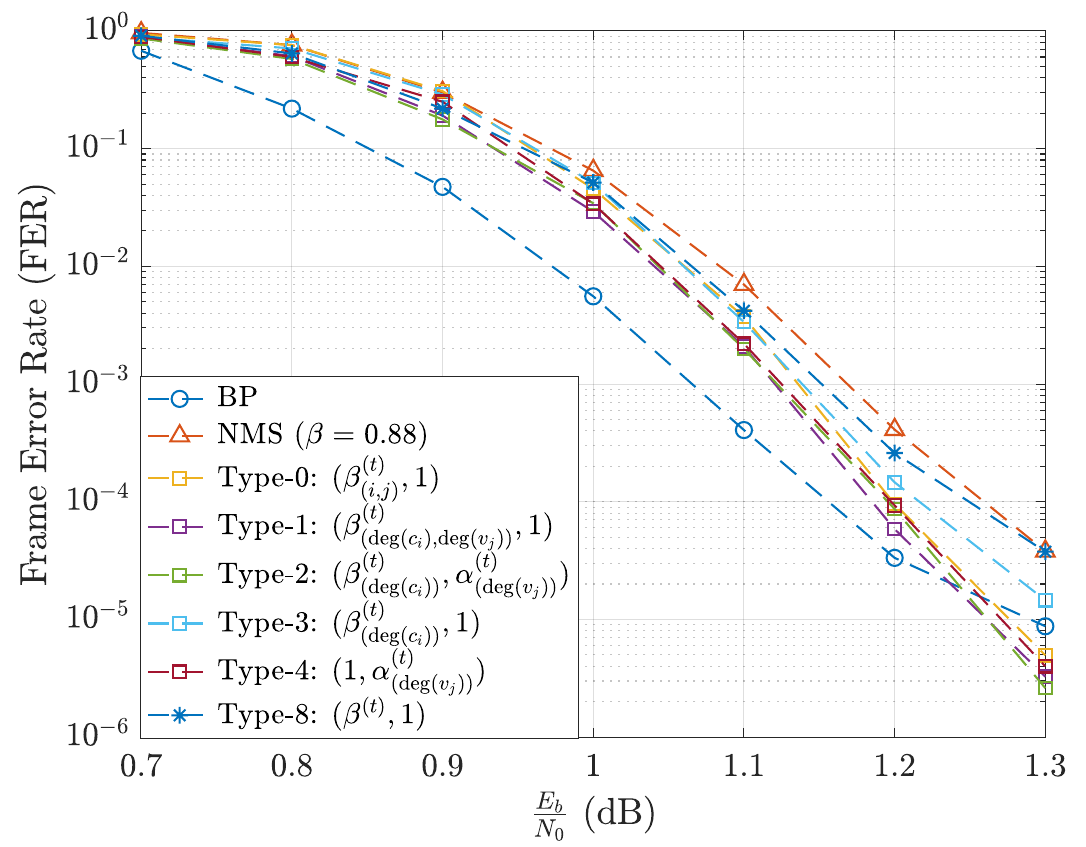}}
  \\
  \subfloat[\label{fig: dvbs2-fer2}]{%
        \includegraphics[width=0.8\linewidth]{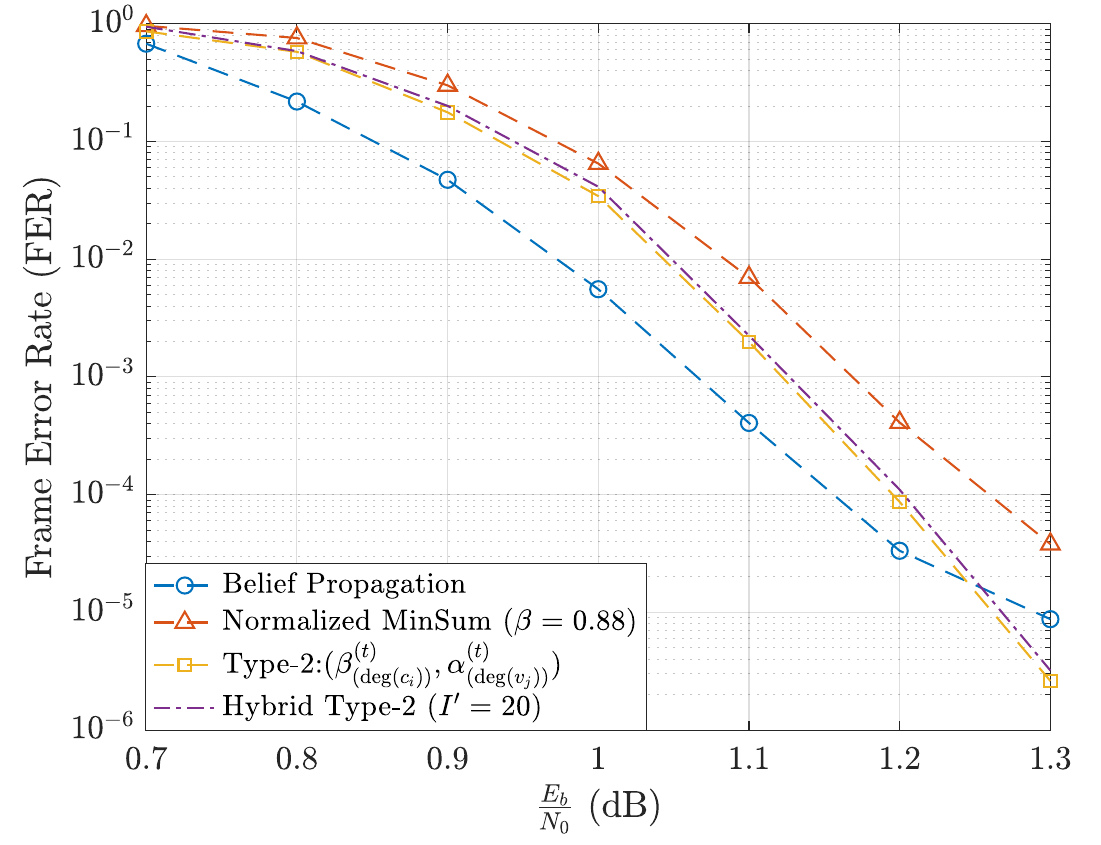}}
  \caption{ Fig. (a): The FER performance of 
  the N-2D-NMS decoders with various weight-sharing types for the (16200,7200) DVBS-2 LDPC code.
  Fig. (b): The FER performance of the hybrid type-2 N-2D-NMS decoder that uses distinct weights in the first 20 iterations and the same weights in the remaining 30 iterations. }
\end{figure}

\subsection{(16200,7200) DVBS-2 LDPC code}
Fig. \ref{fig: dvbs2-fer1} shows the FER performance of N-2D-NMS decoders with various weight sharing types for the (16200, 7200) DVBS-2 LDPC code. 
The FER performance of BP and NMS decoders \textcolor{blue}{is} given for comparison.  The single multiplicative weight of  NMS decoder is 0.88.
All decoders are flooding-scheduled, and the maximum decoding iteration is 50. 
\textcolor{black}{It is shown that the N-NMS decoder (i.e., type-0 decoder) outperforms BP at $1.3$ dB with a lower error floor but requires $4.8*10^{-5}$ parameters in each iteration. The type-1 and type-2 decoders, which share weights based on the check node degree and variable node degree, deliver a slightly better decoding performance than the N-NMS decoder, with only 13 and 8 parameters per iteration, respectively.}

\begin{figure}[t] 
    \centering
 \subfloat[\label{fig: beta_iter}]{%
       \includegraphics[width=0.8\linewidth]{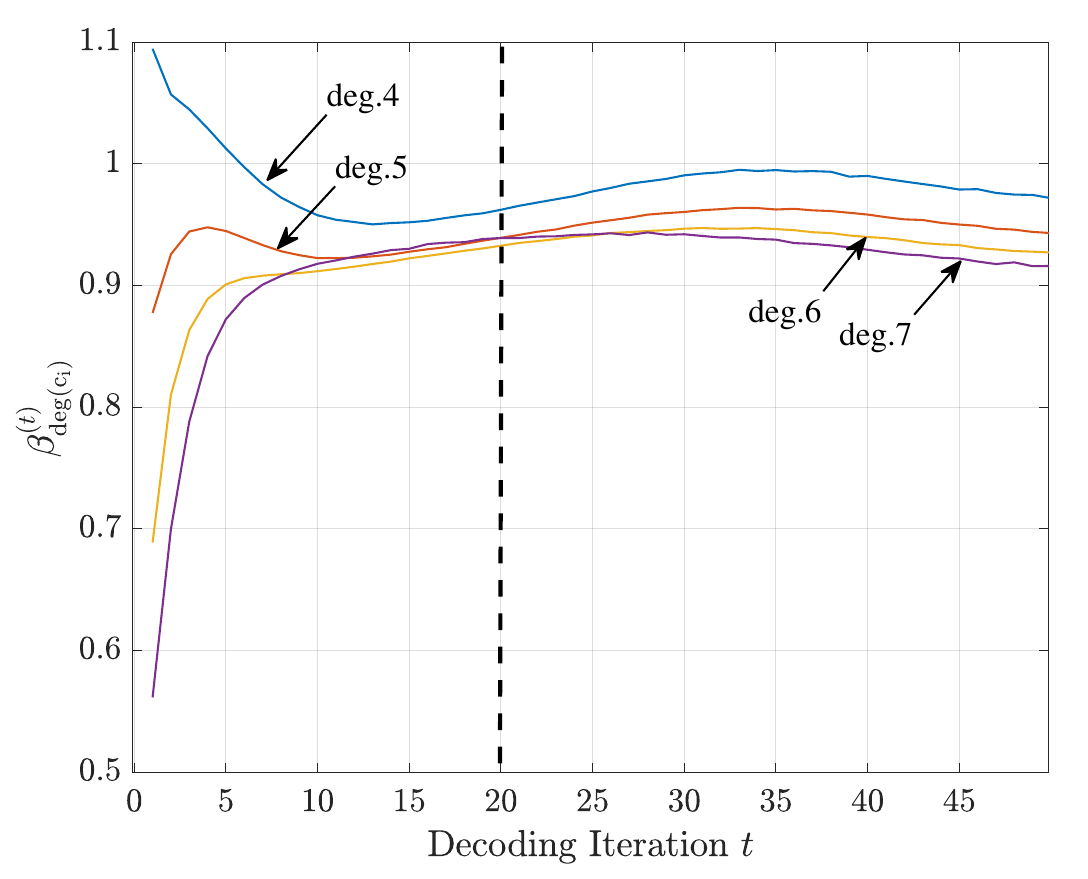}}
 \\
  \subfloat[\label{fig: alpha_iter}]{%
        \includegraphics[width=0.8\linewidth]{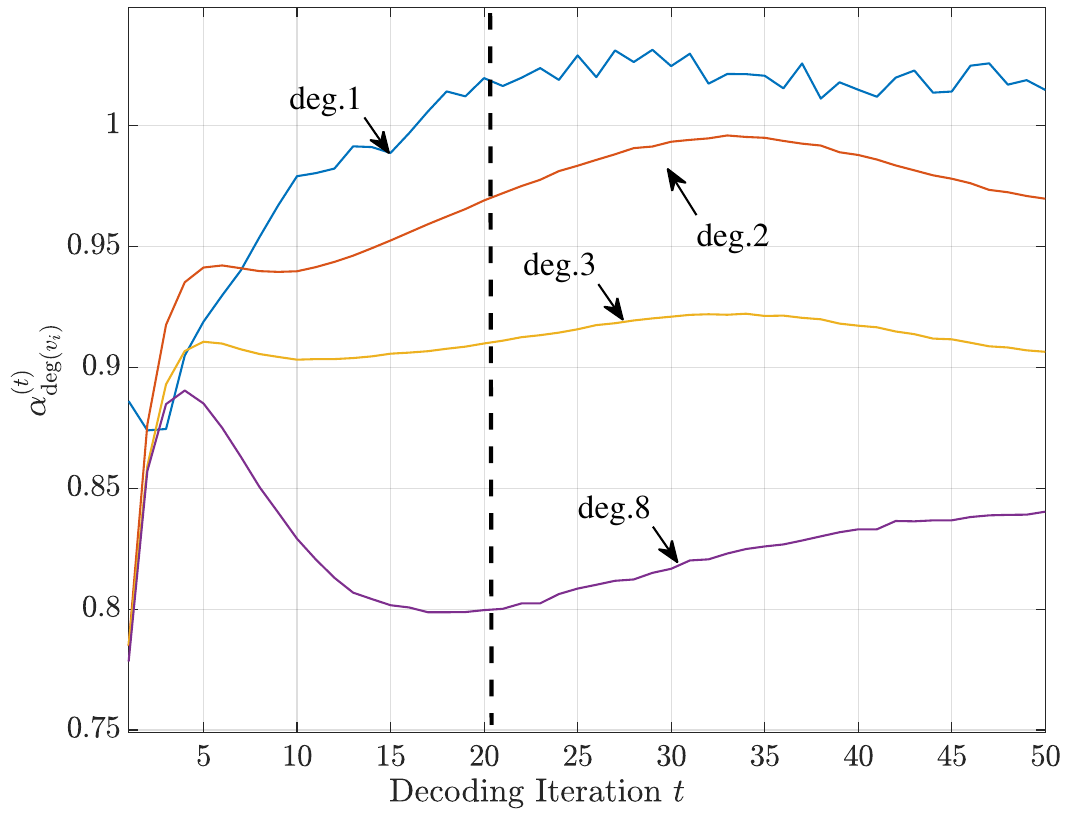}}
  \caption{ The change of weights of the type-2 N-2D-NMS decoder for (16200, 7200) DVBS-2 LDPC code w.r.t. check node degree, variable node degree and iteration index.
  Specifically, Fig. (a) gives $\beta^{(t)}_{(\text{deg}({c_i}))}$ for all possible check node degrees in each decoding iteration $t$, Fig. (b) gives $\alpha^{(t)}_{(\text{deg}({v_j}))}$ for all possible variable node degrees in each decoding iteration $t$.}
\end{figure}

Fig. \ref{fig: dvbs2-fer1} also shows that the FER performance degrades if only considering sharing weights w.r.t. the check node degree (type-3) or the variable node degree (type-4). 
\textcolor{black}{Type-4 decoder delivers a  similar performance to N-NMS decoder; whereas Type-3 decoder is inferior to N-NMS decoder by 0.04 dB. However, both types 3 and 4 require only 4 parameters in each iteration.}
Fig. \ref{fig: beta_iter} and \ref{fig: alpha_iter} give the $\beta^{(t)}_{(\text{deg}({c_i}))}$ and $\alpha^{(t)}_{(\text{deg}({v_j}))}$ of type-2 N-2D-NMS decoder, which \textcolor{blue}{align} with our observation in the previous section; i.e., in each decoding iteration, larger degree node corresponds to a smaller value. Besides, as shown in Fig. \ref{fig: beta_iter} and \ref{fig: alpha_iter}, the weights change negligibly after $20^{th}$ iteration. Thus, the hybrid type-2 N-2D-NMS decoder with $I'=20$ delivers similar performance to the full feed-forward decoding structure, as shown in Fig. \ref{fig: dvbs2-fer2}. 



\subsection{(9472,8192) Quasi-Cyclic LDPC code}
This subsection designs 3-bit and 4-bit W-OMS-RCQ decoders for a (9742,8192) QC LDPC code and compares them with the fixed-point OMS decoder and RCQ decoders. All decoders in this subsection are layer-scheduled with a maximum iteration of 10. 
\textcolor{black}{The 4-bit W-OMS-RCQ decoder uses one quantizer/dequantizer pair with $C=10$, $\gamma=1.7$ in decoding. The  3-bit W-OMS-RCQ decoder, on the other hand, uses three quantizer/dequantizer pairs. The decoder uses the quantizer/dequantizer with $C=3$ and $\gamma=1.3$ in the first six iterations. In iteration 7 to 8, the quantizer/dequantizer has  $C=5$ and $\gamma=1.3$. The quantizer/dequantizer uses  $C=7$ and $\gamma=1.3$ in the last two decoding iterations.}
\textcolor{black}{The pair $(b_c, b_v)$ in the legend gives the bit width of each decoder's check node message and variable node message.}



\begin{figure}[t] 
    \centering
  \subfloat[\label{fig: fer_8k_1}]{%
       \includegraphics[width=0.80\linewidth]{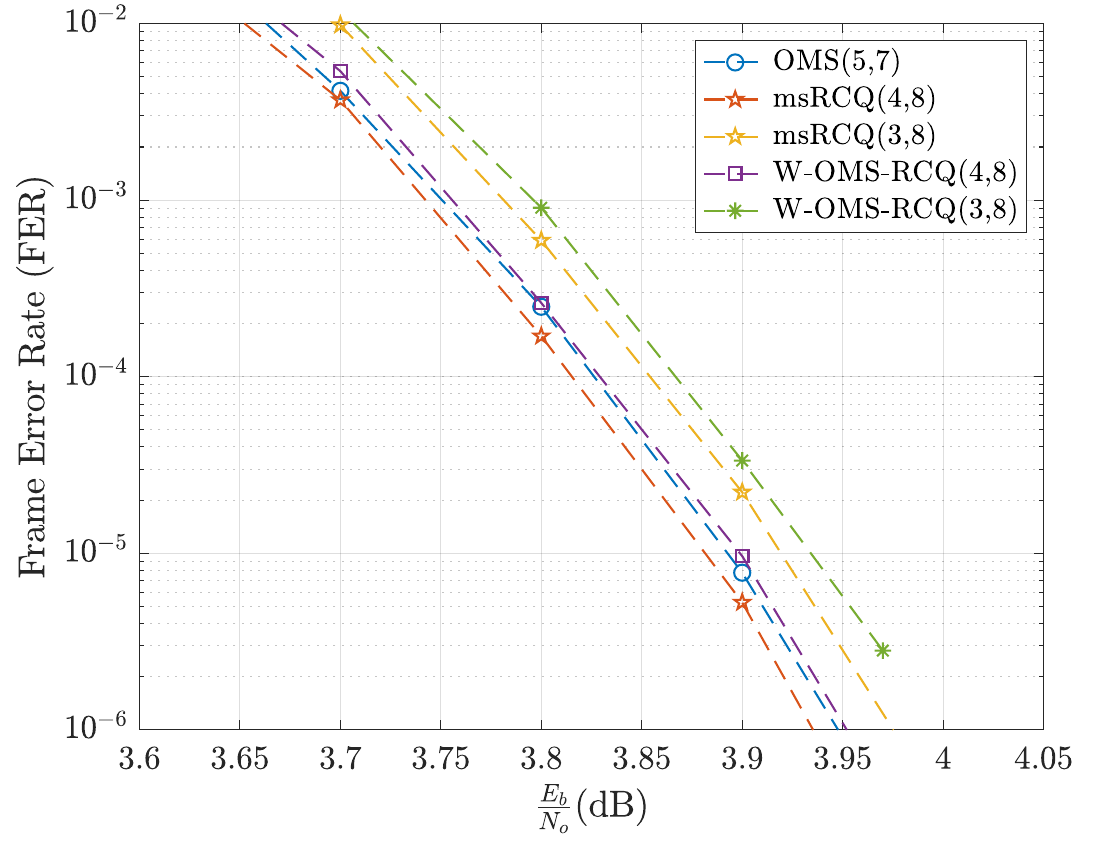}}
   \\
  \subfloat[\label{fig: fer_8k_2}]{%
        \includegraphics[width=0.80\linewidth]{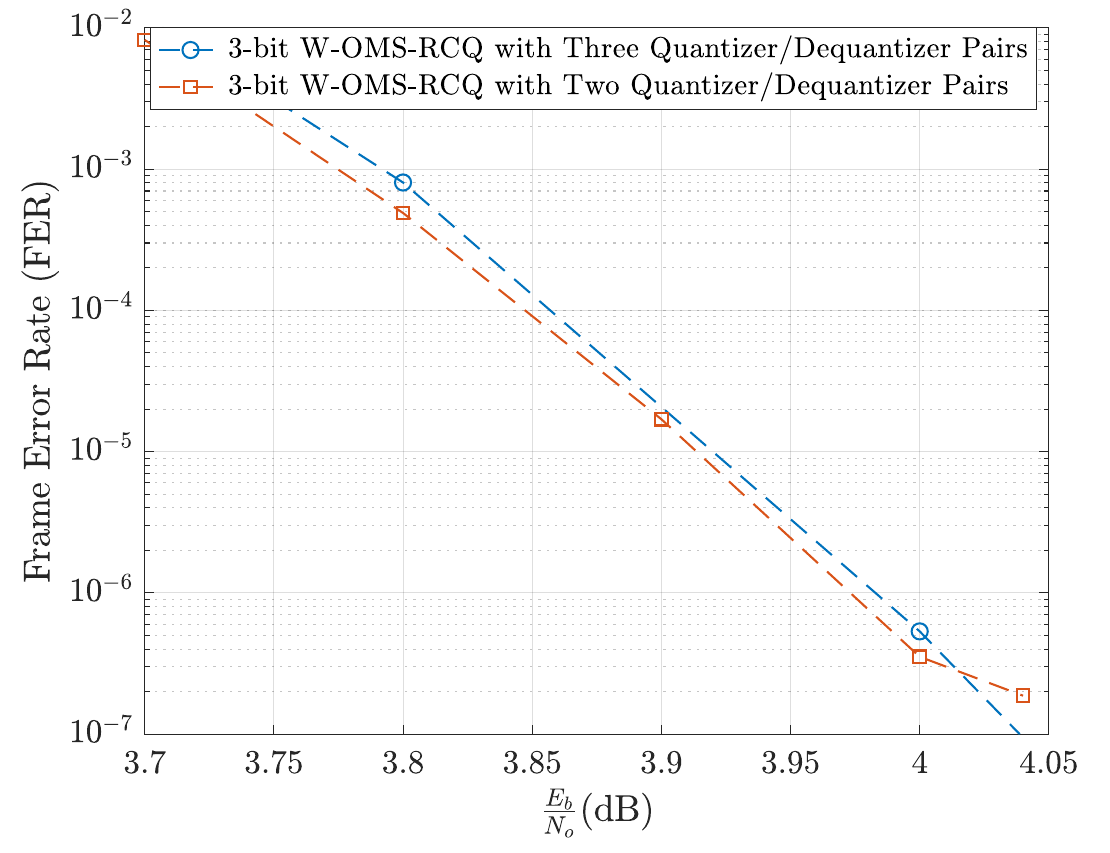}}
  \caption{Fig. (a): FER performance of W-OMS-RCQ decoders, RCQ decoders, and 5-bit OMS decoder for a (9472, 8192) QC LDPC code. Fig. (b): FER performance of 3-bit W-OMS-RCQ decoders with two and three quantizer/dequantizer pairs. }
\end{figure}

Fig. \ref{fig: fer_8k_1} compares the FER performance of W-OMS-RCQ decoders with msRCQ decoders and a 5-bit OMS decoder. The decoders in Fig. \ref{fig: fer_8k_1} are also implemented using an FPGA device (Xilinx Zynq
UltraScale+ MPSoC) for the study of resource usage. Table \ref{tab: hardware_usage} lists the usage of lookup tables (LUTs), registers, block RAM (BRAM), and routed nets of various decoders. For the details of FPGA implementations of the decoders, we refer the readers to \cite{terrill2021fpga}. 

The simulation result shows the 4-bit msRCQ decoder has the best FER performance. The 4-bit W-OMS-RCQ decoder and 5-bit OMS decoder have similar FER performance, which is inferior to the 4-bit msRCQ decoder by 0.01 dB. However, as shown in Table \ref{tab: hardware_usage}, the 4-bit W-OMS-RCQ decoder requires much fewer resources than the 4-bit msRCQ decoder and the 5-bit OMS decoder. 
Compared to the 5-bit OMS decoder, the 3-bit W-OMS-RCQ and 3-bit msRCQ decoder have a 0.025 and 0.05 dB gap, respectively. 
Specifically, the 3-bit msRCQ decoder has similar LUT, BRAM, and routed net usage to the 4-bit W-OMS-RCQ decoder. On the other hand, the 3-bit W-OMS-RCQ uses much fewer resources than the 4-bit W-OMS-RCQ decoder.

\input{hardware_8k}

The 3-bit W-OMS-RCQ decoder in Fig. \ref{fig: fer_8k_1} uses three quantizers for three decoding phases. In the first three iterations, most messages have low magnitudes. Hence a quantizer with small $C$ is required for a finer resolution to the low-magnitude values. However, the message magnitudes increase with the increase of decoding iteration. As a result, the quantizers with larger $C$ should be used correspondingly. Fewer quantizers may not accommodate the message magnitude growth in the decoding process and will result in performance degradation. For example, Fig. \ref{fig: fer_8k_2} considers a 3-bit W-OMS-RCQ decoder that uses two quantizer/dequantizer pairs, the first pair has $C_1=3$, $\gamma_1=1.3$ and is used for iteration $1\sim7$, the second pair has $C_2=5$, $\gamma_2=1.3$ and is used for iteration $8\sim10$. The simulation result shows that the 3-bit W-OMS-RCQ decoder that uses 2 quantizer/dequantizer pairs has an early error floor at FER of $10^{-7}$.
\begin{figure}[t]
	\centering
	\includegraphics[width=20pc]{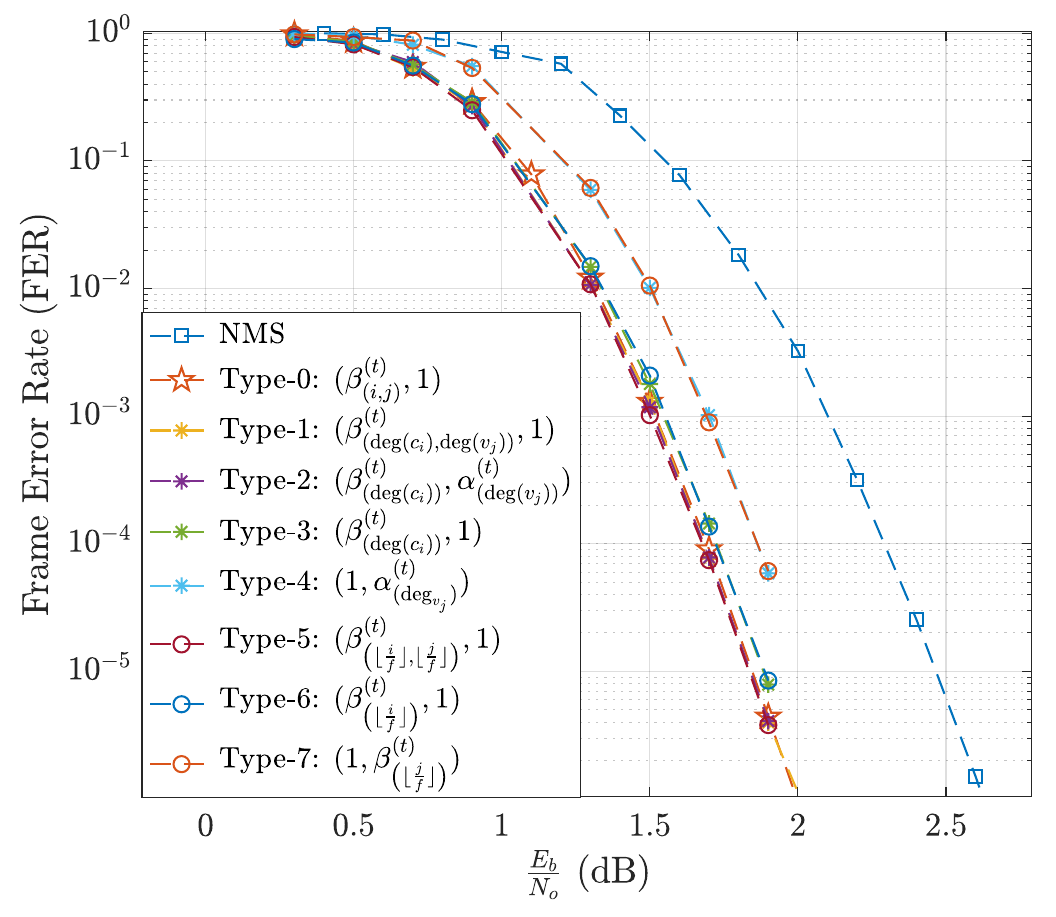}  
	\caption{FER performance of N-2D-NMS decoders with various weight sharing types for a (3096,1032) PBRL LDPC code compared with N-NMS (type 0) and NMS.}
    \label{fig: FER1}
\end{figure}

\subsection{$k=1032$ Protograph-Based Raptor-Like code}

5G LDPC codes have the protograph-based raptor-like (PBRL) \cite{PBRL} structure which offers inherent rate-compatibility and excellent decoding performance. In this subsection, we examine the performance of N-2D-NMS decoders and W-RCQ decoders for a $k=1032$ PBRL LDPC code, whose supported rates are listed in Table \ref{tab:weight_sharing}.  The edge distribution of the lowest-rate code, which corresponds to the full parity check matrix, is also given in Table \ref{tab:weight_sharing}.  All the decoders in this subsection are layer-scheduled with a maximum of 10 decoding iterations.

Fig. \ref{fig: FER1} shows the FER performance of the N-2D-NMS decoder with various weight sharing types for the PBRL code with lowest code rates $\frac{1}{3}$. As a comparison, the decoding performance of the N-NMS (type 0) decoder and the NMS decoder is also given. All of the decoders use floating-point message representation. \textcolor{black}{The simulation results show that the N-NMS decoder has a more than 0.5 dB improvement over the NMS decoder but requires $1.6*10^{4}$ parameters per iteration, as given in Table \ref{tab:weight_sharing}. On the other hand, the N-2D-NMS decoders with types 1, 2, and 5 have the same decoding performance as the N-NMS decoder but only use 41, 15, and 101 parameters in each iteration, respectively. By only considering sharing weights based on check node degrees, N-2D-NMS decoders of types 3 and 6 have a degradation of around 0.05 dB compared with the N-NMS decoder, with 8 and 17 parameters in each iteration, respectively. On the other hand, when only considering sharing the weights based on the variable node degrees, N-2D-NMS decoders of types 4 and 7 have a degradation of around 0.2 dB compared with the N-NMS decoder, with 7 and 25 parameters in each iteration, respectively. Thus, for this (3096,1032) PBRL LDPC code, assigning weights based only on check nodes can benefit more than assigning weights based on variable nodes.}


\begin{figure}[t] 
    \centering
  \subfloat[\label{fig: pbrl_fer_1}]{%
       \includegraphics[width=0.80\linewidth]{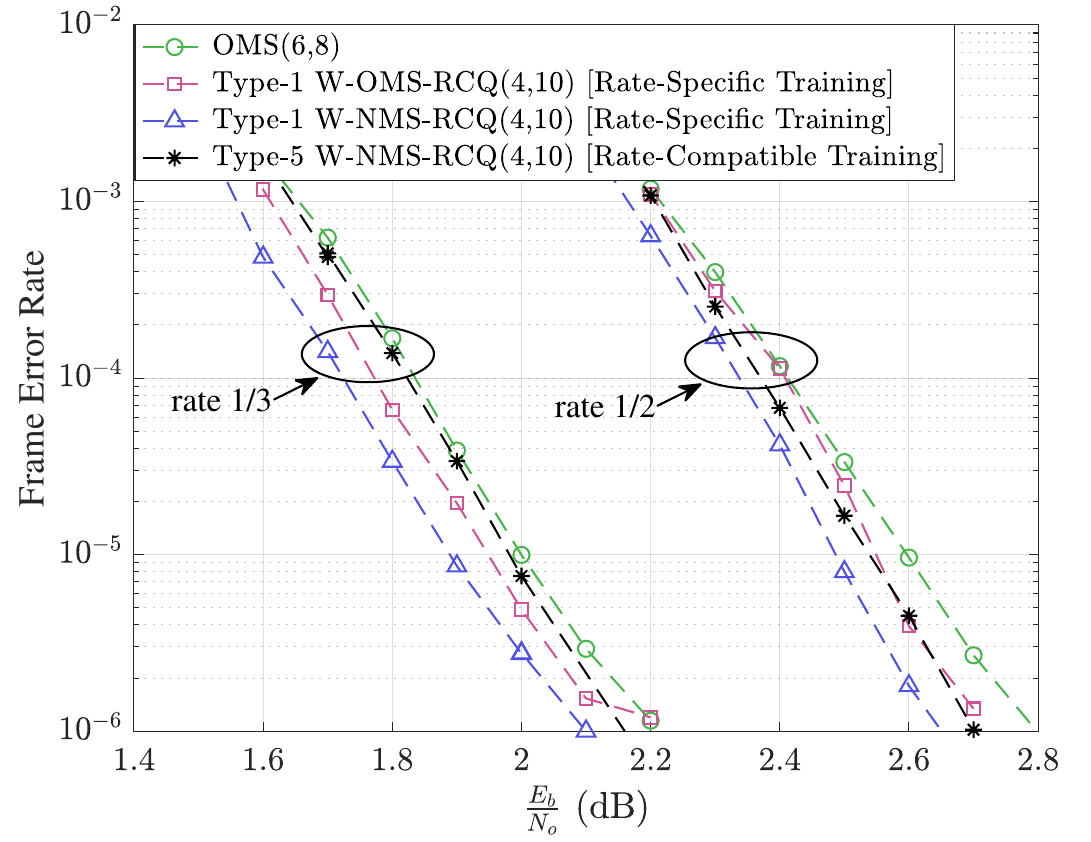}}
\\
  \subfloat[\label{fig: pbrl_fer_2}]{%
        \includegraphics[width=0.80\linewidth]{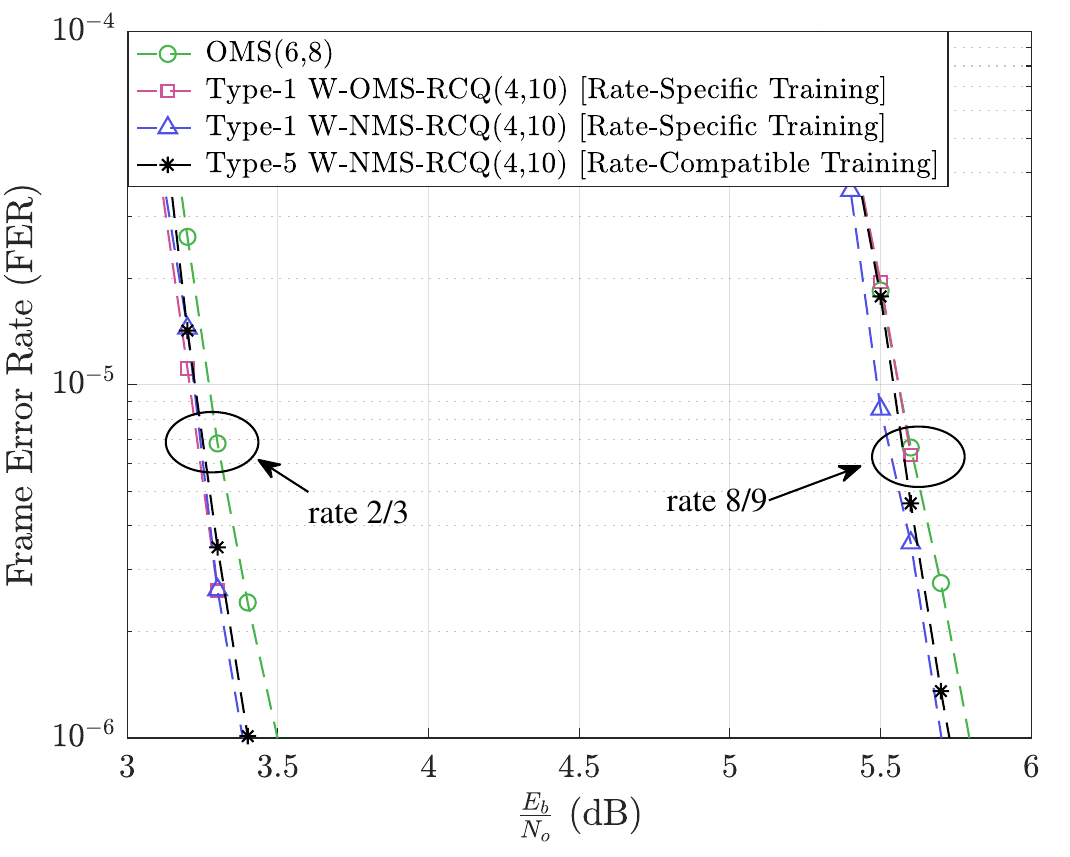}}
  \caption{FER performance of 4-bit W-RCQ decoders for $k=1032$ PBRL code with different code rates. The term "rate-specific" means designing distinct decoders for each code rate; "rate-compatible" means training one decoder that matches all code rates. The 6-bit OMS decoder is given as a comparison. }\label{fig: pbrl}
\end{figure}

Fig. \ref{fig: pbrl} gives the FER performance of fixed-point W-NMS-RCQ decoders for the $k=1032$ PBRL code with rate $\frac{1}{3}$, $\frac{1}{2}$, $\frac{2}{3}$ and $\frac{8}{9}$. The W-NMS-RCQ decoder assigns 4 bits to C2V message and 10 bits to V2C message. Two quantizer/dequantizer pairs are used for W-NMS-RCQ decoder across all investigated rates.
The first quantizer has $C_1=7$, $\gamma_2=1.7$ and is used for the first 7 iterations. The second quantizer has $C_2=10$, $\gamma_2=2.3$ and is used for the last three iterations.
We use a 6-bit OMS decoder as the benchmark because it delivers better decoding performance than the NMS decoder with the same bit width.

We first consider the 4-bit W-NMS-RCQ decoder with type-1 weight sharing that assigns the same weight to the edges with the same check \textcolor{black}{node degree} and variable node degree.
The decoder is rate-specific; i.e., a W-NMS-RCQ decoder is trained separately for each considered code rate.
The simulation results show that, targeting an FER of $10^{-6}$, the 4-bit rate-specific W-NMS-RCQ decoder outperforms the 6-bit OMS decoder with 0.1$\sim$ 0.15 dB for all considered code rates.
\textcolor{black}{
Fig. \ref{fig: pbrl} also gives the FER curves of the 4-bit type-1 rate-specific W-OMS-RCQ decoder at various code rates. The simulation indicates that W-OMS-RCQ doesn't perform as well as the W-NMS-RCQ decoder. }

For the PBRL code, the protomatrix of each possible rate is a sub-matrix of a base protomatrix \cite{PBRL}. As shown in Table. \ref{tab:weight_sharing}, the type-5 weight sharing assigns the same weight to the edges corresponding to the same element in the protomatirx. Hence, it is possible to use \emph{one} trained type-5 neural decoder to match different code rates. We refer to such a decoder as a rate-compatible decoder. In \cite{dai2021learning}, the authors propose training the rate-compatible decoder using samples from different code rates. 

Fig. \ref{fig: pbrl} shows the decoding performance of the rate-compatible type-5 W-NMS-RCQ decoder. The simulation result shows that for the higher rate, such as $\frac{2}{3}$ and $\frac{8}{9}$, the rate-compatible type-5 W-NMS-RCQ decoder has a similar decoding performance to the rate-specific type-1 W-NMS-RCQ decoder.  However, for the lower rates such as $\frac{1}{3}$ and $\frac{1}{2}$, the rate-compatible type-5 W-NMS-RCQ decoder method doesn't deliver decoding performance as well as rate-specific type-1 W-NMS-RCQ decoder.
Besides, considering the four rates in Fig. \ref{fig: pbrl}, the number of neural weights for rate-specific type-1 and rate-compatible type-5  W-NMS-RCQ decoder are 96 and 101, respectively.

\section{Conclusion}\label{sec: conclusion}

Neural networks have improved MinSum message-passing decoders for low-density parity-check (LDPC) codes by multiplying or adding weights to the messages, where a neural network determines the weights.  However, the neural network complexity to determine distinct weights for each edge is high, often limiting the application to relatively short LDPC codes. In particular, when training the neural network using PyTorch or TensorFlow, memory constraints prevent designing weights for long-blocklength codes. This paper solves this memory issue by compactly storing feed-forward messages, which allows us to design weights for blocklengths of 16,000 bits. As an additional contribution, this paper identifies a gradient explosion problem in the neural decoder training and provides a  posterior joint training method that addresses this problem. 

For neural decoders such as N-NMS decoder and N-OMS decoder, assigning distinct weights to each edge in each decoding iteration is impractical for long-blocklength codes because of the storage burden associated with the huge number of the neural weights.
This paper proposes node-degree-based weight-sharing schemes that  drastically reduce the number of required weights with often negligible increase in frame error rate.

Finally, this paper combines the idea of weights designed by a nerual network with the nonlinear quantization paradigm of RCQ, producing  the W-RCQ decoder, a non-uniformly quantized decoder that delivers excellent decoding performance in the low-bitwidth regime.
Unlike the RCQ decoder, which designs quantizer/dequantizer pairs for each layer and iteration, the W-RCQ decoder only uses a small number of quantizer/dequantizer pairs.

\bibliographystyle{IEEEtran}
\bibliography{nn_decoder,csl}

\end{document}

%% file: weight_sharing_one_col.tex

\begin{table}[t]
\centering
\caption{Various Node-Degree-Based Weight Sharing Schemes and\\ Required Number of Parameters per Iteration for Two Example Codes}
\begin{tabular}{|m{7mm}|c|c|c|c|}
\hline
\multirow{2}{*}{Type} & \multirow{2}{*}{$\beta^{(t)}_{*}$}                                    & \multirow{2}{*}{$\alpha^{(t)}_{*}$}                & \multicolumn{2}{c|}{\begin{tabular}[c]{@{}c@{}}The number of Required\\  Parameters per Iteration\end{tabular}}                        \\ \cline{4-5}  
                      &                                                                       &                                                    & \begin{tabular}[c]{@{}c@{}}(16200,7200) \\ DVBS-2 code\end{tabular} & \begin{tabular}[c]{@{}c@{}}(3096,1032) \\ PBRL code\end{tabular} \\ \hline
\multicolumn{5}{|c|}{No Weight Sharing\cite{Nachmani2016-bs}}                                                                                                                                                                                                                                                     \\ \hline
$0$            & $\beta^{(t)}_{(c_i,v_j)}$                                             & 1                                                  & $4.8*10^5$                                                          & $1.60*10^4$                                                      \\ \hline
\multicolumn{5}{|c|}{Weight Sharing Based on Node Degree}                                                                                                                                                                                                                                   \\ \hline
1                     & $\beta^{(t)}_{(\text{deg}{(c_i)},\text{deg}{(v_j)})}$                                 & 1                                                  & 13                                                                  & 41                                                               \\ \hline
2                     & $\beta^{(t)}_{(\text{deg}{(c_i)})}$                                          & $\alpha^{(t)}_{(\text{deg}{(v_j)})}$                      & 8                                                                   & 15                                                               \\ \hline
3                     & $\beta^{(t)}_{(\text{deg}{(c_i)})}$                                          & 1                                                  & 4                                                                   & 8                                                                \\ \hline
4                     & 1                                                                  & $\alpha^{(t)}_{(\text{deg}{(v_j)})}$                      & 4                                                                   & 7                                                                \\ \hline
\multicolumn{5}{|c|}{Weight Sharing Based on Protomatrix}                                                                                                                                                                                                                                   \\ \hline
5\cite{dai2021learning}          & $\beta^{(t)}_{\left( \floor{\frac{i}{f}},\floor{\frac{j}{f}}\right)}$ & 1                                                  & $-$                                                                 & 101                                                              \\ \hline
6                     & $\beta^{(t)}_{\left( \floor{\frac{i}{f}}\right)}$                     & 1                                                  & $-$                                                                 & 17                                                               \\ \hline
7                     & 1                                                                       & $\alpha^{(t)}_{\left( \floor{\frac{j}{f}}\right)}$ & $-$                                                                 & 25                                                               \\ \hline
\multicolumn{5}{|c|}{Weight sharing based on Iteration  \cite {Lian2019-jh,Abotabl2019-wt}}                                                                                                                                                                                                          \\ \hline
8                     & $\beta^{(t)}$                                                         & 1                                                  & 1                                                                   & 1                                                                \\ \hline
\end{tabular}\label{tab:weight_sharing}
\end{table}

%% file: code_info.tex
\begin{table*}[t]
\centering
\caption{LDPC Codes used for Simulation}
\begin{tabular}{|c|c|l|}
\hline
Code & Rate & \multicolumn{1}{c|}{Edge distribution} \\ \hline
\multicolumn{1}{|l|}{(16200,7200) DVBS-2 LDPC code\cite{noauthor_2019-nv}} & $\frac{4}{9}$ & \begin{tabular}[c]{@{}l@{}}$\lambda(x)=2.06*10^{-5}+0.3703x+0.3333x^2+0.2963x^7$\\ $\rho(x)=0.1186x^3+0.3332x^4+0.4445x^5+0.1037x^6$\end{tabular} \\ \hline
(9472,8192) QC-LDPC code \cite{wang2022TCOMRCQ} & $\frac{8}{9}$ & \begin{tabular}[c]{@{}l@{}}$\lambda(x)=x^3$\\ $\rho(x)=0.3919x^{28}+0.6081x^{29}$\end{tabular} \\ \hline
$k=1032$ PBRL LDPC code\cite{cls_tool} & $\frac{8}{9}, \frac{8}{10},\dots,\frac{8}{24}$ & \begin{tabular}[c]{@{}l@{}}$\lambda(x)=0.1190+0.7940x^4+0.0952x^5+0.0556x^6+$\\ $0.3095x^{12}+0.1270x^{16}+0.2143x^{26}$\\ $\rho(x)=0.0238x^2+0.0635x^3+0.0794x^4+0.1905x^5+$\\ $0.2222x^6+0.1270x^7+0.1429x^{17}+0.1508x^{18}$\end{tabular} \\ \hline
\end{tabular}\label{tab: code_info}
\end{table*}

%% file: hardware_8k.tex
\begin{table*}[t]
\caption{\label{tab: hardware_usage} Hardware Usage of Various Decoding Structure for (9472,8192) QC-LDPC Code }
\centering
\begin{tabular}{|c|c|c|c|c|}
\hline
Decoding Structure  & LUTs                                            & Registers                                       & BRAMS                                           & Routed Nets                                     \\ \hline
OMS(5,7) (baseline) & 21127                                           & 12966                                           & 17                                              & 29202                                           \\ \hline
msRCQ(4,8)          & 20355(\textcolor{mygreen}{$\downarrow 3.6\%$} ) & 13967(\textcolor{red}{$\uparrow 7.0\%$})        & 17.5(\textcolor{red}{$\uparrow .03\%$})         & 28916(\textcolor{mygreen}{$\downarrow 1\%$})    \\ \hline
msRCQ(3,8)          & 17865(\textcolor{mygreen}{$\downarrow 15.4\%$}) & 12098(\textcolor{mygreen}{$\downarrow 6.7\%$})  & 17(\textcolor{blue}{$-$})                       & 25332\textcolor{mygreen}{($\downarrow 13.3\%$}) \\ \hline
W-OMS-RCQ(4,8)         & 17645(\textcolor{mygreen}{$\downarrow 16.5\%$} ) & 13297(\textcolor{red}{$\uparrow 2.6\%$}) & 17(\textcolor{blue}{$-$}) & 25361(\textcolor{mygreen}{$\downarrow 13.2\%$} )\\ \hline
W-OMS-RCQ(3,8)         & 16306(\textcolor{mygreen}{$\downarrow 22.82\%$} ) & 12104(\textcolor{mygreen}{$\downarrow 6.65\%$}) & 17(\textcolor{blue}{$-$}) & 23252(\textcolor{mygreen}{$\downarrow 20.38\%$} )\\ \hline
\end{tabular}
\end{table*}